\documentclass[useAMS,usenatbib]{mn2e}

\usepackage[dvips]{graphicx}
\usepackage{natbib}
\usepackage{ulem}
\usepackage{amsmath,amssymb}
\usepackage{mathrsfs}

\usepackage{color}

\newcommand{\msunyr}{{\rm M}_{\odot}~{\rm yr}^{-1}}
\newcommand{\mdot}{\dot{M}}
\newcommand{\msun}{{\rm M}_{\odot}}

\newcommand{\lsun}{{\rm L}_{\odot}}
\newcommand{\zsun}{{\rm Z}_{\odot}}

\newcommand{\araa}{ARA\&A}

\newcommand{\aap}{A\&A}

\newcommand{\apj}{ApJ}
\newcommand{\apjl}{ApJ}
\newcommand{\apjs}{ApJS}

\newcommand{\mnras}{MNRAS}

\newcommand{\nat}{Nature}
\newcommand{\pasj}{PASJ}
\newcommand{\pasp}{PASP}

\newcommand{\na}{Nature}
\newcommand{\ssr}{Space Sci. Rev.}

\title[Gravitational instability in low-metallicity disks]
{Gravitational instability in protostellar disks\\ at low metallicities}
\author[Kei E. I. Tanaka and Kazuyuki Omukai]{
Kei E. I. Tanaka$^{1,2}$
\thanks{E-mail: ktanaka@astr.tohoku.ac.jp}
and Kazuyuki Omukai$^{1,2}$
\thanks{E-mail: omukai@astr.tohoku.ac.jp}
\\
$^{1}$Astronomical Institute, Tohoku University, Sendai 980-8578, Japan\\
$^{2}$Department of Physics, Kyoto University, Kyoto 606-8502, Japan\\
}
\begin{document}


\pagerange{\pageref{firstpage}--\pageref{lastpage}} \pubyear{2002}

\maketitle

\label{firstpage}
\begin{abstract}
Fragmentation of protostellar disks controls the growth of protostars and 
plays a key role in determining the final mass of newborn stars.
In this paper, we investigate the structure and gravitational stability 
of the protostellar disks in the full metallicity range between zero and the solar value.
Using the mass-accretion rates evaluated from the thermal evolution in the preceding 
collapse phase of the pre-stellar cores, we calculate disk structures and 
their evolution 
in the framework of the standard steady disks.
Overall, with higher metallicity, more efficient cooling results 
in the lower accretion rate and lower temperature inside the disk: 
at zero metallicity, the accretion rate is $\sim 10^{-3}\msunyr$ and the disk
temperature is $\sim 1000~{\rm K}$, while at solar metallicity, 
$\sim 10^{-6}\msunyr$ and $\sim 10~{\rm K}$.
Despite the large difference in these values,
the zero- and solar-metallicity disks have similar stability properties:
the Toomre parameter for the gravitational stability, which can be written 
using the ratio of temperatures in the disk and in the envelope 
as $Q_{\rm T} \sim (T_{\rm disk}/T_{\rm env})^{3/2}$ , is  
$\ga 1$, i.e., marginally stable.
At intermediate metallicities of $10^{-5}$ -- $10^{-3}\zsun$, however,
the disks are found to be strongly unstable with $Q_{\rm T}\sim 0.1-1$
since dust cooling, which is effective only in the disks 
due to their high density ($\ga10^{10}{\rm cm}^{-3}$), 
makes the temperature in the disks lower than that in the envelopes.
This indicates that masses of the individual stars formed as a result of 
the protostellar disk fragmentation can be significantly smaller than 
their parent core in this metallicity range. 
The typical stellar mass in this case would be a few $\msun$,
which is consistent with the observationally suggested mass-scale of extremely metal-poor stars.
\end{abstract}

\begin{keywords}
early universe - stars: formation - stars: Population II - accretion, accretion disks.
\end{keywords}

\section{Introduction}
Stars in the early universe are considered to have played important roles 
in setting the environment for subsequent star formation in young galaxies, 
and starting the reionization and metal enrichment of the intergalactic medium 
through their radiation and kinetic energy injection in the supernovae
\citep[e.g.,][]{cia05}.
Since the degree of those feedbacks strongly depends on the mass of stars,
a number of studies have been carried out to pin down their mass range.
Over the last decade, the first star formation from primordial pristine gas 
has been investigated in great depth, and their typical mass is found to be 
$10$ -- $100\msun$ \citep{mck08,hos11,hir13}.
On the contrary,
the initial mass function of present-day stars is observationally known to peak 
at $\la 1\msun$ in the solar neighborhood \citep{kro02,cha03}.
These facts indicate the existence of a transition of the stellar mass-scale 
during the cosmic history.

A promising hypothesis is that this mass-scale transition has been caused by 
the fragmentation of pre-stellar clouds owing to the cooling by accumulated metals
\citep{omu00,bro01,sch02,bro03,omu05,omu10,omu12}.
In particular, for fragmentation into sub-solar mass clumps, 
rapid cooling at some high density, where the Jeans mass is small enough, is required. 
The cooling by dust thermal emission is considered to have played this role. 
Analytic studies \citep{sch03,sch06,omu05,omu10} as well as 
numerical simulations \citep{cla08,dop11,dop13}
demonstrated that star-forming clouds enriched with metallicity above a critical value 
$\sim 10^{-5}\zsun$ fragment at high density $\ga 10^{10}{\rm cm^{-3}}$
by the dust cooling 
in the case of the same dust properties
(i.e., depletion factor, size distribution, composition, etc.) 
as in the solar neighborhood.
The critical metallicity remains at similar value even for models of the dust 
produced in the first-star supernovae although with some uncertainty, e.g., 
due to destruction by supernova reverse shocks, 
growth by coagulation during the pre-stellar collapse, etc. 
\citep{sch06,sch12,noz12,chi13}

So far, those studies have mainly focused on the fragmentation in the pre-stellar collapse, 
i.e., the preceding phase to the birth of protostars.  
The protostars, however, acquire most of their mass through disk accretion
in the later so-called main-accretion phase.
Low-mass clumps can thus be formed also by fragmentation of protostellar disks.
In fact, even in the pristine-gas case,
about $\sim 1/2$ of the protostellar disks are found to fragment during the 
main-accretion phase, bearing low-mass clumps, according to numerical simulations
\citep{sta10,sta13}.
With increasing metallicity and thus enhanced dust cooling, 
the temperature in the disk will decrease.
If the disk temperature becomes so low that the thermal pressure cannot 
cope with the disk self-gravity anymore, the disk will fragment into a number of smaller 
clumps. 
In the stellar cluster formed in this way, each member would be typically far less 
massive than their parent cloud.
Here, to see the fragmentation properties of protostellar disks at low metallicites,
we study their self-gravitational stability using the standard steady 
thin-disk prescription.

Some previous works have discussed the stability of low-metallicity disks
with planet formation in low-metallicity environments in mind, 
and have concluded that they can be even more unstable than 
in the solar-metallicity disks
\citep[][see also Boss 2002]{cai06,mer10}.
However, the metallicity range studied has been limited to rather high values of 
$10^{-2}\la Z/\zsun \la 10$, relevant to observed exoplanets.
No comprehensive study has been carried out covering the entire metallicity range 
between zero and solar value.
In addition, in those studies the initial disk structures were set up arbitrarily 
by hand. In reality, disk properties, such as the accretion rate and 
disk radius, depend on the metallicity of star forming gas, 
which determines the condition of accreting envelope.  
In this paper, we calculate the evolution of protostellar disks 
considering the accretion histories set by the structure of their parental clouds 
for the entire range of metallicities $0\leq Z \leq \zsun$.

This paper is organized as follows.
In Section 2, we describe the standard steady disk model adopted in this study.
In Section 3, to extract the effect of different metallicity on the disk stability,
we present the disk structures for given accretion rates and disk radii.
Next, in Sectin 4, we illustrate evolution of the disks under accretion rates 
set by the pre-stellar collapse with various metallicites.
In Section 5, we discuss roles of disk fragmentation 
on setting the final mass of stars formed and uncertainties in our model.
Finally, we summarize our study in Section 6.

\section{Steady disk model} \label{sec_disk_model}
We here describe our model for the protostellar disks.
Although the protostellar disks, especially in their early phase, 
evolve dynamically both in zero- \citep{sta10,cla11,vor13}
and the solar-metallicity cases \citep{wal09,vor10,mac10,tsu11}, 
we adopt the conventional steady-state $\alpha$-disk prescription \citep{sha73}
to explore disk properties in a wide range of parameters 
(metallicity, stellar mass, disk radius, and accretion rate).
We calculate the disk structure considering the gravity only from its central star,
and then discuss its stability against the disk self-gravity.
We use the Toomre parameter for the self-gravitational stability \citep{too94},
\begin{eqnarray}
	{Q}_{\rm T} = \frac{c_{\rm s}\kappa_{\rm ep}}{\pi G\Sigma}, \label{eq_Q}
\end{eqnarray}
where
$\kappa_{\rm ep}$ is the epicycle frequency,
$\Sigma$ the disk surface density,
$c_{\rm s}=\sqrt{k_{\rm B} T/\mu m_{\rm H}}$ the sound speed,
$T$ the temperature at the midplane,
$\mu$ the mean molecular weight,
$k_{\rm B}$ the Boltzmann constant,
$m_{\rm H}$ the proton mass, and $G$ the gravitational constant.
The epicycle frequency is given by the Keplerian angular velocity,
$\kappa_{\rm ep} = \Omega_{\rm Kep} \equiv \sqrt{GM_*/r^3}$,
where $M_*$ is the stellar mass and $r$ is the orbital radius.
The disk structure with $Q_{\rm T}<1$ is unstable for self-gravity 
and will fragment into clumps.

Let us consider the density and thermal structure of a disk 
with accretion rate $\dot{M}$.
The density structure is determined by hydrostatic equilibrium 
in the vertical direction, and the mass and angular momentum conservation.
From the vertical hydrostatic equilibrium, $c_{\rm s}^2/H=GM_*H/r^3$,
the disk scale hight $H$ at the radius $r$ is 
\begin{eqnarray}
	H = \frac{c_{\rm s}}{\Omega_{\rm Kep}}. \label{eq_H}
\end{eqnarray}
In the innermost region of $r\la1{\rm AU}$,
the temperature can exceed $T \ga10^5{\rm K}$ and 
the radiation pressure can be important \citep{tan11}.
We here, however, neglect the contribution of radiation pressure
since we are interested in the disk structure and stability in the larger scale. 
The disk surface density $\Sigma$ is evaluated from the assumption of steady accretion,
\begin{eqnarray}
	\Sigma = \frac{\dot{M}}{3\pi \nu}, \label{eq_steady}
\end{eqnarray}
where $\nu$ is the kinematic viscosity \citep[e.g.,][]{pri81}.

The thermal structure of the disks is determined by the thermal equilibrium 
at each radius
\begin{eqnarray}
	{\mathscr H} = {\mathscr L}, \label{eq_th_eq}
\end{eqnarray}
where ${\mathscr H}$ and ${\mathscr L}$ are heating and cooling rate, respectively, 
per unit surface area.
We consider the heating owing to the turbulent viscosity: 
\begin{eqnarray}
	{\mathscr H} = \frac{9}{4}  \nu \Sigma \Omega_{\rm Kep}^2. \label{eq_heat}
\end{eqnarray}
The heating by the stellar radiation
is not included in the calculation and its influence will be discussed 
later in Section \ref{sec_uncertain}.
As cooling processes, we consider radiative cooling by the $\rm {H}_2$ lines 
and the dust and gas continuum.
Using the cooling rate per unit volume $\Lambda$, the surface cooling rate 
can be written as 
\begin{eqnarray}
	{\mathscr L} = 2H \Lambda .
\end{eqnarray}
The ${\rm H}_2$-line cooling rate is calculated as in \citet{omu05} by 
solving rovibrational level populations for given H$_2$ column density.
The most hydrogen is already in the molecular form in the disks due to 
the high density (typically $\ga 10^{10} {\rm cm}^{-3}$). 
We just assume that the gas is fully molecular and do not solve the chemical reaction equations.
Although the cooling by ${\rm H_2O,~HD, {\rm~and~}CO}$ lines  
as well as by fine-structure lines of ${\rm [C_I],~[C_{II}],{\rm~and~}[O_I]}$ 
can be important in some density and metallicity ranges in the pre-stellar collapse phase,
the dominant coolant in the high density range as in protostellar disks 
is always either $\rm H_{2}$ or the continuum \citep{omu05,omu10}.
The continuum cooling rate by the gas and dust is calculated by using 
the opacities for the gas $\kappa_{\rm g,i}$ and dust $\kappa_{\rm d,i}$, 
where the subscript {\rm i}={\rm P,R} means the Planck and Rosseland mean, respectively:
\begin{eqnarray}
	&&\Lambda_{\rm cont} = \frac{4 \sigma_{\rm SB} \rho}{1+x}
		\left(  \kappa_{\rm g, P} T^4 + \kappa_{\rm d, P} T_{\rm d}^4 \right),
	\label{eq_cont}  \\
	&&x = \tau_{\rm P} + 3\tau_{\rm R}\tau_{\rm P}/4,\\
	&&\tau_{i} = \left( \kappa_{{\rm g},i} + \kappa_{{\rm d},i} \right) \Sigma,
	~~~~~~~ i={\rm P,R}.
\end{eqnarray}
where
$\sigma_{\rm SB}$ is the Stefan-Boltzmann constant,
$\rho = \Sigma/(2H)$ the density,
$T_{\rm d}$ the dust temperature,
and $\tau_{\rm P, R}$ the optical depth.
The function $x(\tau_{\rm i})$ gives the correct limiting values both in 
the optically thin and thick regimes and smoothly connects them.
For the gas opacity, we use that for the primordial gas by \citet{may05}.
We use the dust opacity by \citet{sem03} in the solar metallicity case, 
which is reduced in proportion to the metallicity in other cases.

The dust temperature $T_{\rm d}$ is calculated from the energy balance between 
the dust thermal emission and energy transfer by collisions with gas particles:
\begin{eqnarray}
	4\sigma \rho  \kappa_{\rm d,P} 
	\left( T_{\rm d}^4 - T_{\rm rad}^4 \right)
	=
	\frac{n_{\rm d} \left( 2k_{\rm B} T - 2k_{\rm B}T_{\rm d} \right)}
	{t_{\rm coll}} \label{eq_dust_eq},
\end{eqnarray}
where $T_{\rm rad}$ is the temperature of radiation field inside the disk, 
$n_{\rm d}$ the dust number density, and $t_{\rm coll}$ the mean free time 
between the collisions
\citep{hol79,sch06}.
The radiation temperature is given by
\begin{eqnarray}
	T_{\rm rad}^4 = \frac{x}{1+x} 
	\frac{\kappa_{\rm g, P} T^4 + \kappa_{\rm d, P} T_{\rm d}^4}
	{\kappa_{\rm g, P} + \kappa_{\rm d, P}},
	\label{eq_Trad}
\end{eqnarray}
\citep[See Appendix B4 of][for derivation]{omu01}.
The factor $x/(1+x)$ on right-hand side is valid both in the optically thin and thick 
limits and connects them smoothly. 
Note that, in the energy-balance equation (eq. \ref{eq_dust_eq}), 
both the dust opacity $\kappa_{\rm d,P}$ on the right-hand side and 
the dust number density $n_{\rm d}$ on the left-hand side being proportional 
to the metallicity,
the dust temperature is independent of the metallicity.
The dust and gas thermally couple above the density 
\begin{eqnarray}
	n_{\rm H, tc} \simeq 2.3 \times 10^{11} \left( \frac{T}{100{\rm K}} \right)^{4.5} {\rm cm}^{-3}
	\label{eq_ncr}
\end{eqnarray}
(see Appendix \ref{sec_nc} for the derivation),
while the dust temperature is much lower than the gas temperature 
below $n_{\rm H, tc}$.

We treat the transport of angular momentum by means of viscosity with the $\alpha$-parameter 
\begin{eqnarray}
	\nu = \alpha c_{\rm s} H, \label{eq_alpha}
\end{eqnarray}
\citep{sha73}.
We divide $\alpha$ into two components, those by the gravitational instability (GI) and by the magnetic rotational instability (MRI): $\alpha = \alpha_{\rm GI} + \alpha_{\rm MRI}$.
It is known that the torque due to the gravitational instability is significant in protostellar disks \citep{bat98,kru09,cla11,kui11}.
The gravitational torque is efficient when the disk is marginally stable with $Q_{\rm T}\la1.5$.
This efficient torque prevents the disk from becoming too unstable.
Once the Toomre parameter becomes less than unity by efficient cooling or high surface density, the disk fragments.
In order to mimic such behavior of the gravitational-torque efficiency,
$\alpha_{\rm GI}$ is often treated as a functional form of $Q_{\rm T}$ \citep{lin87, nak94,nak95,kra08, ric09, zhu09, tak13}.
We also apply the following formula with which $\alpha_{\rm GI}$ increases abruptly around $Q_{\rm T}=1.5$ and saturates at maximum value of at around $Q_{\rm T}=1$,
\begin{eqnarray}
	\alpha_{\rm GI} = \alpha_{\rm GI,max} \exp\left(-Q_{\rm T}^{10}/10 \right),
	\label{eq_agi}
\end{eqnarray}
where $\alpha_{\rm GI,max}$ is the maximum value of $\alpha_{\rm GI}$.
Various star formation simulations showed that $\alpha$ can be as large as $0.1$ -- $1$ via the gravitational torque \citep{kru07,kra10,cla11,kui11}.
Thus, we set $\alpha_{\rm GI,max}=1$ as our reference value.
However, the exact value of maximum $\alpha$ is still controversial.
Numerical simulations of isolated disks without infalling flow from the envelope suggested smaller maximum $\alpha$ of $0.07$ \citep{gam01,ric05}.
While the star formation simulations follow every stages in star formation from the cloud cores, the isolated disk simulations are well controlled for extracting the fundamental physics of fragmentation.
Our study cannot tell which maximum $\alpha$ value is more feasible.
Therefore, we also show the disk stability adopting the maximum $\alpha$ of $0.07$ to see the dependence on its value.
When the disk is gravitationally stable ($Q_{\rm T}\ga1.5$), the MRI viscosity dominates the angular momentum transfer, for which we apply a fixed value of $\alpha_{\rm MRI}=0.01$ \citep{dav10,shi10}.
We should note that, in reality, $\alpha_{\rm MRI}$ varies depending on the ionization degree of the disk gas.
Especially, in the ``dead zone" where the ionization degree is too low to activate MRI,
$\alpha_{\rm MRI}$ would be zero \citep{gam96,san00,bai11}.
However, the choice of the $\alpha_{\rm MRI}$ value does not affect our conclusion,
since the protostellar disks we are interested in are usually marginally or strongly unstable with $Q_{\rm T}\la1.5$.
We will discuss the uncertainty of $\alpha_{\rm GI}$ and $\alpha_{\rm MRI}$ and its influences on our conclusion in Section \ref{sec_uncertain}.

We should note that our steady-disk framework is unable to describe the dynamics of fragmentation process and its outcome.
We apply the pseudo-viscosity $\alpha=\alpha(Q_{\rm T})$, which corresponds to the strength of turbulence.
In fact, numerical simulations by \citet{cos09} showed that the amplitude of turbulence in a self-gravitating disk is a function of the cooling rate rather than the Toomre parameter: the turbulent amplitude is regulated to produce heating rate that balances the cooling rate to maintain the disk in a marginally stable state.
If the cooling rate is too high, the required turbulent amplitude enters into the non-linear regime.
Then, the Toomre parameter drops locally $Q_{\rm T}<1$ and the disk fragments.
This underlying physics of fragmentation is not strictly captured by our model,
which treats the turbulence by the viscous parameter $\alpha(Q_{\rm T})$.
However, our model is still appropriate to study the disk stability by the following reasons.
First, the viscous heating is consistently regulated to balance the cooling (eq. \ref{eq_th_eq}).
Second, our $\alpha(Q_{\rm T})$ prescription is chosen in order to mimic that fragmentation  occurs when the turbulent amplitude becomes too large:
$Q_{\rm T}<1$ once $\alpha$ reaches the maximum (eq. \ref{eq_agi}).
Additionally, local fluctuations of $Q_{\rm T}$ in a self-gravitating disk are relatively small and thus the average $Q_{\rm T}$ is also almost close to unity when the disk fragments.
Because of these facts,
the obtained solution gives the reasonable description of the disk structure
even with the subtle difference from the realistic fragmentation process.
Note that, although a disk with $Q_{\rm T}<1$ is expected to fragment, we will present the disk structure also in the cases with $Q_{\rm T}<1$ as a guide for understanding the disk stability.

\section{Local disk structures for given radii and accretion rates} \label{sec_local}
Both the disk radius and accretion rate depend on structure of 
the parent pre-stellar core and thus on the metallicity. 
In this section, however, to see how the dust cooling affects 
the disk structure and stability,
we present the results for given radii $r$ and accretion rates $\dot{M}$.
For the solar metallicity case, similar studies have been carried out by \citet{cla09} and \citet{cos10}.

\begin{figure*}
\begin{center}
\includegraphics[width=170mm]{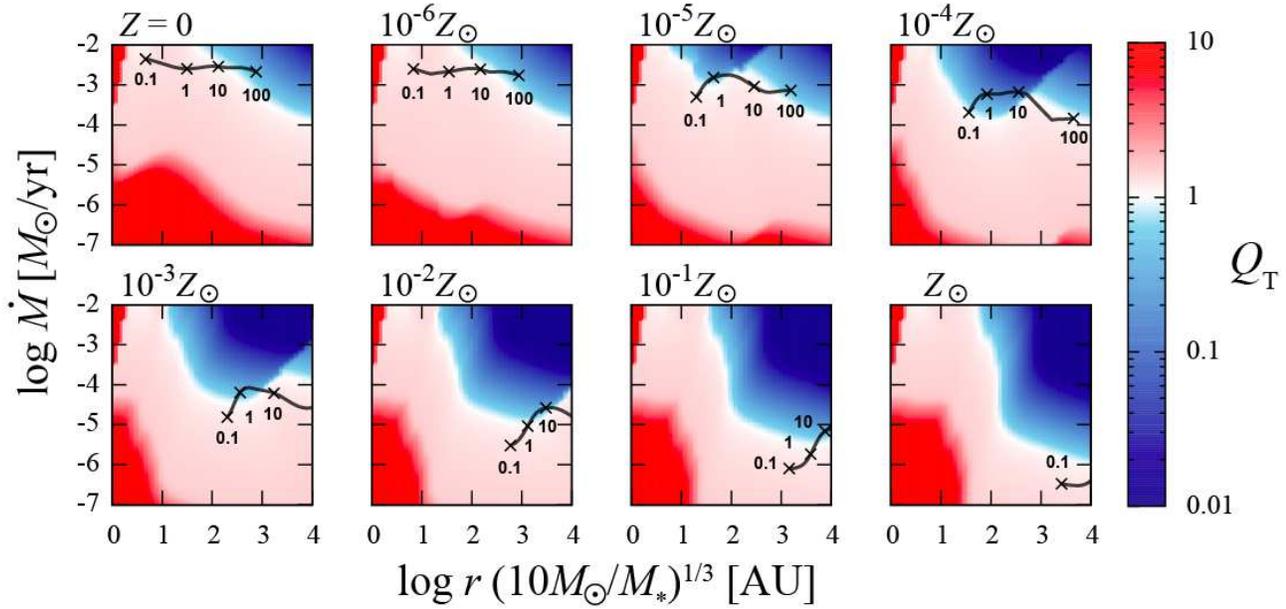}
\end{center}
\caption{
Gravitational stability of the disks with metallicities $Z/\zsun=0,~10^{-6},~10^{-5},~10^{-4},~10^{-3},~10^{-2},~10^{-1},{\rm~and~} 1$
(from top-left to bottom-right)
in the case of $\alpha_{\rm GI,max}=1$. 
In each panel, the Toomre parameter $Q_{\rm T}$ is indicated by color contours 
for given radii (horizontal axis) and accretion rates (vertical axis).
Note that the reduced radius $r(10\msun/M_*)^{1/3}$ is used on the horizontal 
axis (see text in Sec. 3).
Curves illustrate evolutionary tracks of 
the outer disk radii $r_{\rm d}$ and accretion rates $\dot{M}$ 
for the central protostellar mass $M_{\rm *}=0.1$ -- $100\msun$, 
and the crosses on them correspond to $M_{\rm *}=0.1,~1,~10,{\rm~and~}100\msun$ 
(see text in Sec. \ref{sec_result2}).
}
\label{fig_r-Mdot}
\end{figure*}

Figures \ref{fig_r-Mdot} illustrates the Toomre parameter $Q_{\rm T}$ 
on the $r$-$\dot{M}$ planes for different metallicities
with $\alpha_{\rm GI,max}=1$.
Since $M_*$ and $r$ appear in equations for disk structure 
(eq. \ref{eq_Q}, \ref{eq_H}, and \ref{eq_heat}) only through 
$\Omega_{\rm Kep}=\sqrt{GM_*/r^3}$,
we here introduce the reduced radius $r(10\msun/M_*)^{1/3}$ and use it as 
the horizontal-axis variable.
Curves with cross symbols indicate evolutionary tracks of the outer 
disk radii and the accretion rates for the protostellar disks, 
which will be discussed in Section \ref{sec_result2}.
We can see that disks are marginally stable ($1<Q\la1.5$)
for wide area of parameter space at every metallicity.
This is due to self-regulation process in a self-gravitating accretion disk:
if the disk becomes marginally unstable with $Q_{\rm T}\simeq1.5$,
the turbulence boosts the viscous heating rate and the efficiency of the angular momentum transfer (i.e., through enhanced $\alpha$ value in our prescription), 
which works to keep the disk from becoming too unstable \citep{bat98,gam01,tak13}.
It can also recognized for all of the metallicities
the Toomre parameter $Q_{\rm T}$ tends to be small in the upper-right regions in all panels, 
i.e., disks are more unstable with higher $\dot{M}$ or at larger $r$.
From the following expression for the Toomre parameter (eq. \ref{eq_Q}), 
\begin{eqnarray}
	Q_{\rm T} = \frac{3\alpha c_{\rm s}^3}{G\dot{M}}
	\simeq 4.8 \alpha
	\left( \frac{T}{1000 {\rm K}} \right)^{3/2}
	\left( \frac{\dot{M}}{10^{-3} \msun {\rm yr}^{-1}} \right)^{-1}.
	\label{eq_q_cs_Mdot}
\end{eqnarray}
where equations (\ref{eq_H}) -- (\ref{eq_alpha}) have been used, we immediately 
see that, with higher accretion rate, the disk is more massive and thus 
more unstable.
The dependence of $Q_{\rm T}$ on the radius $r$ is through the disk temperature 
$T$, which is lower at larger $r$, since the gravitational potential is 
shallower and so the viscous heating rate is smaller there.
In the innermost region of several AU, on the other hand,
the disk remains stable even with accretion rate as high as 
$\ga 10^{-3}\msun {\rm yr}^{-1}$ owing to very high temperature ($\ga10^4{\rm K}$).
With the increase of metallicity, enhanced dust-cooling rate makes the disk 
temperature lower and thus the unstable region extends toward lower accretion rate.
Note that, in the $10^{-5}$ and $10^{-4}\zsun$ cases with $\ga10^{-4}\msunyr$,
the disk is more unstable in a region around $100{\rm AU}$ (in terms of the reduced radius) than in the outer part around $1000{\rm AU}$. 
This is because the dense environment in the inner disk is needed to activate the dust cooling in such a low-metallicity environment.
With more metals of $10^{-3}$ -- $1\zsun$,
the disk in the region of $10$ -- $100{\rm AU}$ tends to be stabilized.
This is because the disk is optically thick to the dust absorption and
becomes stable owing to the resultant high temperature
\citep{cla09,cos10}.

\begin{figure*}
\begin{center}
\includegraphics[width=170mm]{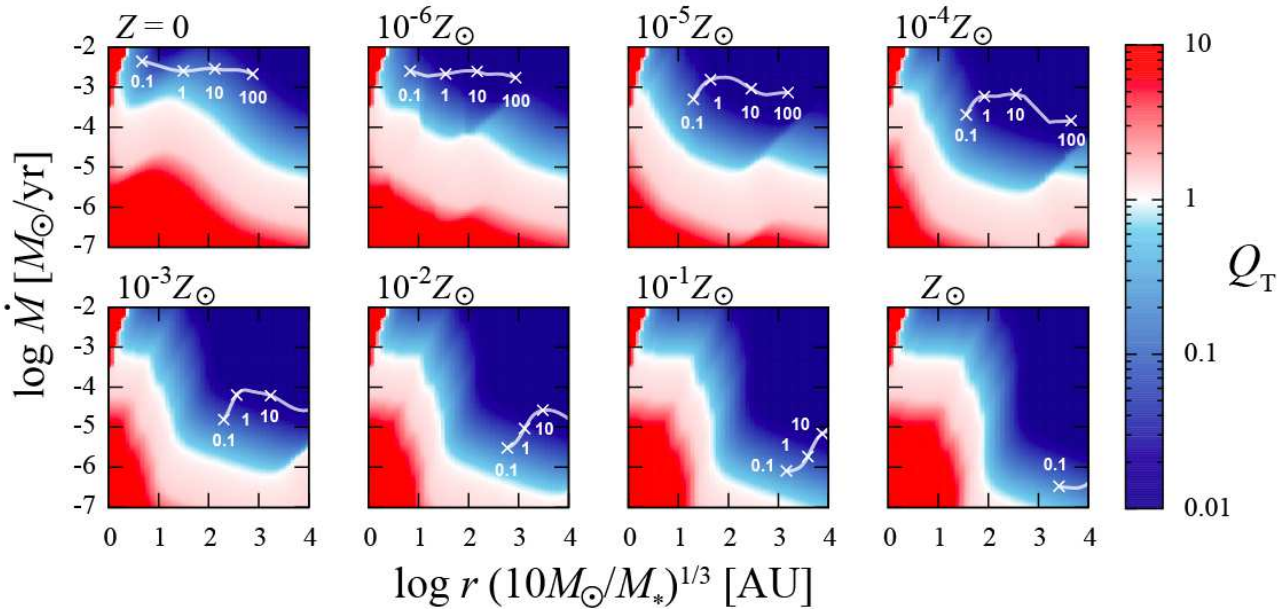}
\end{center}
\caption{
Same as Figure \ref{fig_r-Mdot} except for the maximum $\alpha$ of $0.07$
($\alpha_{\rm GI,max}=0.06$ and $\alpha_{\rm MRI}=0.01$).
}
\label{fig_r-Mdot2}
\end{figure*}

In Figure \ref{fig_r-Mdot2},
we present the disk stability at various metallicities adopting the maximum $\alpha$ of $0.07$, which is suggested by isolated disk simulations \citep{gam01,ric05}.
At the solar metallicity, the fragmentation boundary agrees with those shown by \citet{cla09} and \citet{cos10}, who adopted the maximum $\alpha \simeq0.06$ -- $0.09$.
Comparing with the cases of $\alpha_{\rm GI,max}=1$,
which is indicated from star formation simulations \citep{kru07,kra10,cla11,kui11},
the marginally unstable region shrinks and the unstable region broadens.
This is because the self-regulation effect by the gravitational torque is weaker for smaller maximum $\alpha$-value.
As seen in equation (\ref{eq_q_cs_Mdot}), if temperature is constant,
the critical accretion rate for $Q_{\rm T}=1$ is inversely proportional to the maximum value of $\alpha$.
In this way, the maximum $\alpha$ value determines the efficiency of self-regulation and thus the fragmentation boundary of $Q_{\rm T}=1$.
However, the general trends are same in both cases:
disks are more unstable with higher accretion rates and larger radii for every metallicity,
and unstable region extends toward lower accretion rate as the increase of metallicity and dust cooling efficiency.
Therefore, we will show mainly the results with $\alpha_{\rm GI,max}=1$ hereafter
and discuss the influence of the uncertainty of $\alpha$ value in Section \ref{sec_uncertain}.

\begin{figure}
\begin{center}
\includegraphics[width=75mm]{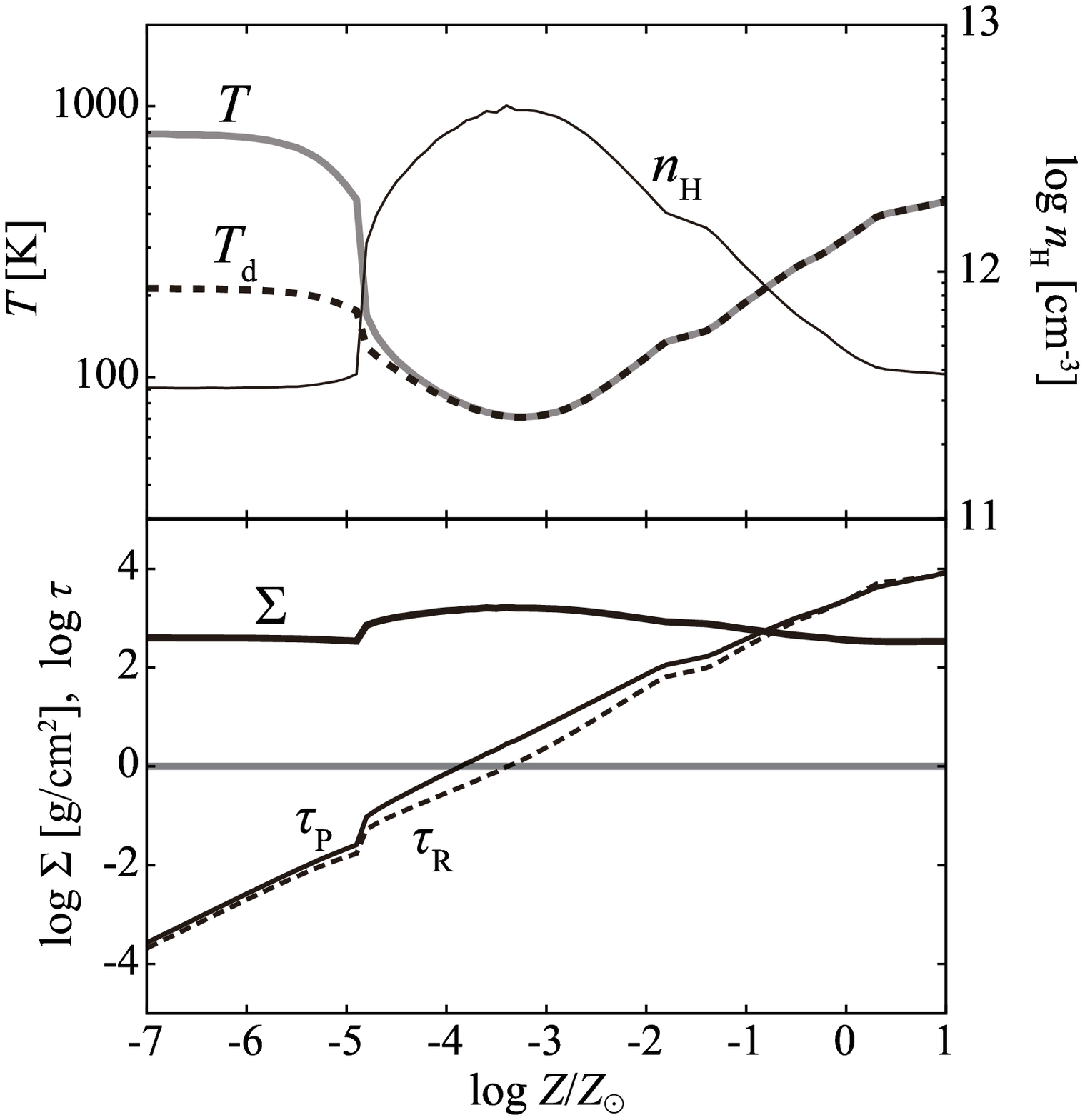}
\end{center}
\caption{Metallicity dependence of local disk structure in the reference case 
where $M_{\ast}=10M_{\odot}, \dot{M}=10^{-3}M_{\odot}{\rm yr^{-1}}$, and $r=100{\rm AU}$.
{\it Top}: the gas temperature $T$, the dust temperature $T_{\rm d}$, and the number density $n_{\rm H}$.
{\it Bottom}: the surface density $\Sigma$, the Planck- ($\tau_{\rm P}$) and Rosseland-mean ($\tau_{\rm R}$) 
optical depths. The gray horizontal line in the bottom panel represents the optically thick/thin boundary of 
$\tau=1$.
The gas temperature reaches the minimum around $\tau=1$.
We should note that the structure for $Z\simeq10^{-5}$ -- $1\zsun$ is unstable ($Q_{\rm T}<1$; see Fig. \ref{fig_Z-Q}) and our steady disk framework is not able to describe such an unstable structure self-consistently.
}
\label{fig_Z-T_n_tau}
\end{figure}

We see the disk structure in the reference case, where 
$M_*=10\msun$, $\dot{M}=10^{-3}\msunyr$, and $r=100{\rm AU}$ 
and its metallicity dependence in more detail.
In both the primordial star formation \citep{sta86, omu98,abe02,bro04,yos06}
and coincidentally in massive star formation in the local universe \citep[e.g.][]{zin07}, 
the typical accretion rate is $\sim 10^{-3}\msunyr$ 
and the disk radius is about $\sim 100{\rm AU}$ at $10\msun$ \citep{tan04}.
Figure \ref{fig_Z-T_n_tau} shows
the number density $n_{\rm H}$, the gas and dust temperatures 
$T$, and $T_{\rm d}$ ({\it upper panel}),
and the surface density $\Sigma$, the Planck and Rosseland-mean optical depths $\tau_{\rm P}$ and $\tau_{\rm R}$
({\it lower panel}) in the reference case as functions of metallicity.
Below metallicity $10^{-5}\zsun$, where the ${\rm H}_2$ line 
dominates the cooling, the gas temperature is constant at $\sim 1000{\rm K}$, 
while the dust temperature remains about $200{\rm K}$, lower than the gas temperature.
With metallicity exceeding $10^{-5}\zsun$,
the dust thermal emission begins to dominate the cooling.
The gas and dust thermally couple each other and have similar 
temperatures, which decrease with increasing metallicity until 
$Z \simeq 10^{-3}\zsun$.
At this point, the optical depth reaches unity (bottom panel of 
Fig. \ref{fig_Z-T_n_tau}).  
Note that the optical depth is roughly proportional to the metallicity
since the surface density 
is relatively constant within an order of magnitude difference 
for different metallicities
(bottom panel of Fig. \ref{fig_Z-T_n_tau}).
With higher metallicity $Z\ga 10^{-3}\zsun$, 
the disk becomes optically thick and the radiative cooling becomes inefficient. 
Now the temperature turns up with metallicity. 
Even with the same $M_*, \dot{M},{\rm~and~}r$, 
the gas temperature varies with an order of magnitude depending on the metallicity.
As seen in the upper panel of Figure \ref{fig_Z-T_n_tau}, 
the number density $n_{\rm H}$ and the gas temperature $T$ behave 
in the opposite way in response to varying metallicity. 
This can be understood as follows: 
the lower temperature leads the smaller scale height (by eq. \ref{eq_H})
and the higher surface density due to smaller viscosity 
(by eq. \ref{eq_steady} and \ref{eq_alpha}; bottom panel of Fig. \ref{fig_Z-T_n_tau}),
and thus the higher density.
Note that the density is always higher than $10^{11}{\rm cm}^{-3}$,
which justifies our assumption of the fully molecular gas in this parameter range.

\begin{figure}
\begin{center}
\includegraphics[width=75mm]{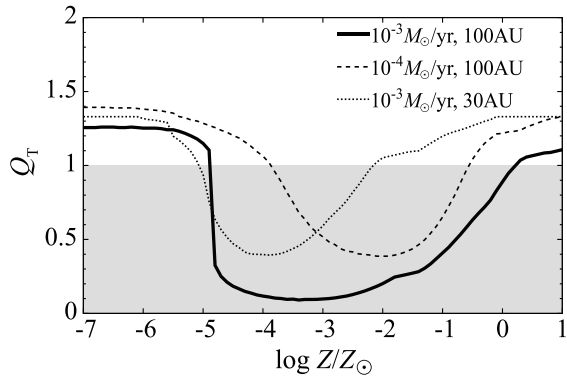}
\end{center}
\caption{The Toomre parameter $Q_{\rm T}$ as a function of metallicity 
for the reference and two other cases for comparison.
All cases are with $M_*=10\msun$.
The accretion rates and disk radii are:
(i) $10^{-3}\msun {\rm yr}^{-1}$ and $100{\rm AU}$ 
({it solid}; the reference case),
(ii) $10^{-4}\msun {\rm yr}^{-1}$ and $100{\rm AU}$ ({\it dashed}),
and 
(iii) $10^{-3}\msun {\rm yr}^{-1}$ and $30{\rm AU}$ ({\it dotted}).
Gray area indicates the instability domain against self-gravity ($Q_{\rm T}<1$).
}
\label{fig_Z-Q}
\end{figure}

Figure \ref{fig_Z-Q} shows the metallicity dependence of the Toomre 
parameter $Q_{\rm T}$ in three cases including the reference one.
From comparison between $Q_{\rm T}$ and $T$ (top panel of Fig. \ref{fig_Z-T_n_tau})
in the reference case ($10^{-3}\msun {\rm yr}^{-1}$ and $100{\rm AU}$), 
we find their metallicity dependences are very similar because 
$Q_{\rm T} \propto T^{3/2}$ (eq. \ref{eq_q_cs_Mdot}). 
Like $T$, $Q_{\rm T}$ also reaches the minimum $\simeq 0.1$ at $\tau \simeq1$.
In reality, such an unstable disk would fragment and 
could not continue accretion steadily. 
In Figure \ref{fig_Z-Q}, two other cases are also shown: 
one with smaller accretion rate ($10^{-4}\msun {\rm yr}^{-1}$)
and the other with smaller radius (30AU) than in the reference case.
In all cases, $Q_{\rm T}$ values take the minimum 
around $10^{-4}$ -- $10^{-2}\zsun$, although their quantitative values are different.
In the case with lower accretion rate $10^{-4}\msun {\rm yr}^{-1}$,
the disk is less massive and less unstable than 
in the reference case ($10^{-3}\msun {\rm yr}$)
as indicated by equation (14). 
In this case, $Q_{\rm T}$ takes its minimum at somewhat higher metallicity of 
$10^{-2}\zsun$ than in the reference case ($\sim 10^{-3}\zsun$)
since the disk becomes optically thick only at higher metallicity 
due to the lower surface density.
Also, at the more inner radius $30{\rm AU}$,
the disk is less unstable than in the reference case ($100{\rm AU}$)
because the temperature is higher there due to the deeper gravitational potential.
The minimum $Q_{\rm T}$ is attained at lower metallicity ($\sim 10^{-4}\zsun$)
than in the reference case because of the higher surface density at the smaller radius. 

\section{Global disk structure and its evolution} \label{sec_global}
In the previous section,
we present the results for a wide range of the accretion rate 
$\dot{M}$ and radius $r$.
In realistic star-forming environments, 
both the accretion rate and the disk size depend on the 
properties of inflows from the envelope.
The envelope structure is set up during the gravitational collapse of 
the parent pre-stellar core and depends on the metallicity.
In this section, we construct models for infalling envelopes 
and calculate the protostellar-disk structures for given metallicities.

\subsection{Model for infalling envelope} \label{sec_env}
As in the standard scenario for present-day star formation,
we here consider the collapse of a pre-stellar core 
from slightly gravitationally unstable state \citep{lar69,shu77,sta80}.
We also assume the core is spherically symmetric for simplicity 
since centrifugal force is not important in the region 
far outside the disk, which we are currently intrested.  
The pre-stellar core undergoes so-called the runaway collapse:
only the central densest part, which 
becomes smaller and smaller in mass, collapses significantly 
leaving outer less dense material almost unevolved.
The size of the central part is roughly given by the instantaneous 
Jeans length and its collapse timescale is about the free-fall time.
Eventually, a low-mass ($\sim0.01\msun$) protostar is formed 
at the center while most of the gas is left behind in the surrounding envelope.
Because the density and temperature in the envelope remain almost unchanged
after that portion of the gas is detached from the central part 
until the protostar formation at the center, 
the envelope structure can be constructed from the thermal evolution in 
the center part during the collapse.
\citet{omu05} investigated
thermal evolution of the pre-stellar clouds during the collapse
by way of a one-zone model, considering detailed chemical and 
radiative processes.
This model provides us with the sound speed $c_{\rm s, 1z}$ 
as a function of density $\rho_{\rm s, 1z}$ for each metallicity. 
The radial mass distribution in the envelope is constructed 
assuming that inside the radius of the Jeans length  
$R_{\rm env} = c_{\rm s, 1z} \sqrt{ {\pi}/{G\rho_{\rm 1z}}}$, 
a Jeans mass of gas $M_{\rm env} = \rho_{\rm 1z} R_{\rm env}^3$
is contained \citep{hos09}.
After the protostar formation at the center, 
the gas in the envelope starts accreting onto it.
The infall rate from the envelope $\dot{M}_{\rm env}$ can be estimated from
the enclosed mass $M_{\rm env}$
divided by the free-fall time 
$t_{\rm ff}$, and given by
\begin{eqnarray}
\dot{M}_{\rm env} \simeq \frac{M_{\rm env}}{t_{\rm ff}} 
\simeq \frac{c_{\rm s, 1z}^3}{G}, \label{eq_infall}
\end{eqnarray}
\citep{shu77}.
We regard this infall as hitting onto the outer disk edge 
and the accretion rate through the disk is constant of radius, i.e., 
$\dot{M}=\dot{M}_{\rm env}$, from the steady assumption. 
Note that the mass in the disk $M_{\rm d} \simeq \pi r_{\rm d}^2 \Sigma(r_{\rm d})$ 
is always small in comparison with the stellar mass:
from equations (\ref{eq_Q}) and (\ref{eq_H}), their ratio can be written 
as $M_{\rm d}/M_* \simeq H/(r_{\rm d}Q_{\rm T})$, which 
indicates that $M_{\rm d}/M_* \simeq H/r \ll 1$
even in the case of a massive disk with $Q_{\rm T}=1$.
Neglecting the mass in the disk, 
we set the instantaneous protostellar mass $M_*$ equal 
to the total accreted mass $M_{\rm env}$.

\begin{figure}
\begin{center}
\includegraphics[width=75mm]{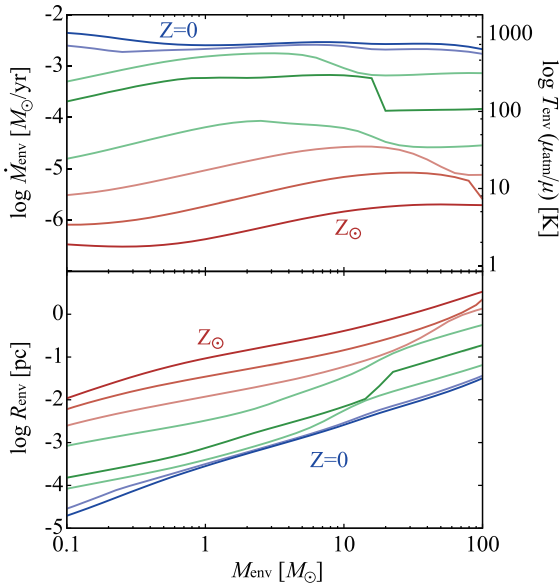}
\end{center}
\caption{The envelope structures at various metallicities.
The infall rate $\dot{M}_{\rm env}$ ({\it top}) and the radius $R_{\rm env}$ 
({\it bottom}) are plotted as functions of the enclosed mass $M_{\rm env}$.
In the top (bottom) panel, $Z/\zsun = 0$, $10^{-6}$ (blue), 
$10^{-5},~10^{-4}, 10^{-3}$ (green), $10^{-2},~10^{-1},~{\rm and}~1$ (red) 
from top to bottom (from bottom to top, respectively).
The envelope temperature $T_{\rm env}$ divided by 
the molecular weight $\mu$ is indicated on the right vertical axis of the top panel. 
Here, the molecular weight is normalized with its value for the fully 
atomic gas $\mu_{\rm atm}\simeq 1.2$.}
\label{fig_envelope}
\end{figure}

Figure \ref{fig_envelope} shows the infall rate $\dot{M}_{\rm env}$ and 
the radius $R_{\rm env}$ as functions of the enclosed mass $M_{\rm env}$ 
for various metallicities.
Recall that the infall rate depends only on the sound speed of the envelope gas,
$\mdot_{\rm env} \propto {c_{\rm s, 1z}^3} \propto (T_{\rm env}/\mu)^{3/2}$
(eq. \ref{eq_infall}).
On the right vertical axis of top panel in Figure \ref{fig_envelope},
$T_{\rm env}/\mu$ is indicated, where $\mu$ is normalized by its value for 
the fully atomic gas $\simeq 1.2$. 
Note the gas is almost atomic in the outer envelope with $M_{\rm env} \ga$ a few $\msun$, 
while it is mostly molecular more inside with $\mu \simeq2.3$.
With metallicity as low as $10^{-6}\zsun$, 
the envelope structure deviates little from that of zero metallicity.
With higher metallicity, the temperature in the envelope and 
so the infall rate become lower 
due to the cooling either by dust ($\ga 10^{-5}\zsun$) or 
by metal lines ($\ga 10^{-4}\zsun$) \citep{omu05,omu10}.
The decrease of infall rate with metallicity can be 
roughly fitted as $\dot{M}_{\rm env} \sim 10^{-3} \left( Z/10^{-6}\zsun \right)^{-0.5} 
\msun {\rm yr}^{-1}$.
The higher temperature at lower metallicity results in the more compact 
infalling envelope: the envelope radius 
$R_{\rm env}\sim GM_{\rm env}/c_{\rm s,env}^2$ 
is smaller for the same enclosed mass.

We have constructed the envelope structure under the assumption of 
the spherical symmetry.
In reality, however, materials in the envelope have non-zero 
angular momentum.  
As a result, a protostellar disk is eventually formed and the accretion proceeds 
onto the protostar through it.
To evaluate the disk radius, we need to model the angular momentum 
distribution in the envelope.
We parameterize the envelope rotation using the ratio of rotational velocity 
$v_{\rm rot,env}$ to the Keplerian velocity 
$v_{\rm Kep,env}=\sqrt{GM_{\rm env}/R_{\rm env}}$:
\begin{eqnarray}
f_{\rm Kep} = \frac{v_{\rm rot,env}}{v_{\rm Kep,env}}
\end{eqnarray}
and assume it has a constant value in the envelope.
In studies on the first star formation, this parameter has been found
$f_{\rm Kep} \simeq 0.5$ without significant dependence on 
radius \citep{abe02,yos06}.
We here adopt $f_{\rm Kep} = 0.5$ as the fiducial value, but 
also study a case of smaller value $f_{\rm Kep} = 0.25$ 
to see the effect of difference in rotation degree.
The conservation of angular momentum in the infall 
leads to the disk outer radius 
\begin{eqnarray}
	r_{\rm d}
	= f_{\rm Kep}^2 R_{\rm env}. \label{eq_fkep}
\end{eqnarray}

\subsection{Results} \label{sec_result2}

\begin{figure*}
\begin{center}
\includegraphics[width=170mm]{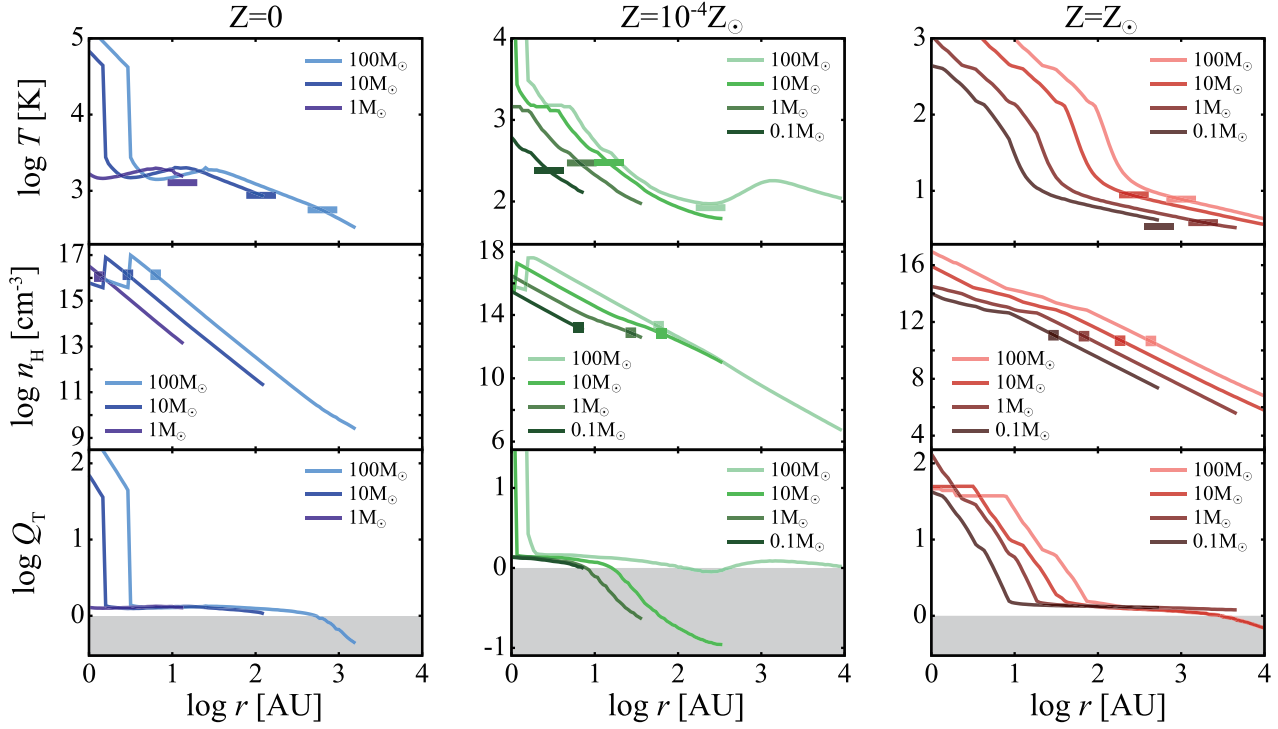}
\end{center}
\caption{The radial structures of the protostellar disks 
and their evolution for metallicities 
$Z/\zsun=0,~10^{-4},{\rm ~and~}1$ (from left to right) 
with the fiducial rotation parameter $f_{\rm Kep}=0.5$.
The distribution of temperature $T$, density $n_{\rm H}$, and Toomre parameter $Q_{\rm T}$
are shown at protostellar masses of $M_*=0.1,~1,~10$, and $100\msun$
(in the $Z=0$ panel, the line for $0.1\msun$ is not shown
because the disk outer radius is smaller than $1{\rm AU}$).
Thick bars in top panels indicate the envelope temperatures $T_{\rm env}$ in the  
corresponding epochs.
Square symbols in middle panels show the radii where the Planck-mean optical depth is equal to unity.
Gray areas in bottom panels represent the unstable region with $Q_{\rm T}<1$.
}
\label{fig_r-TQ}
\end{figure*}

Figure \ref{fig_r-TQ} shows the protostellar-disk structure 
at four epochs $M_*=0.1,~1,~10,{\rm~and~}100\msun$
for $Z/\zsun=0,~10^{-4},~{\rm and}~1$
with the fiducial rotation parameter $f_{\rm Kep}=0.5$:
top, middle, and bottom panels present the radial distributions of the temperature $T(r)$,
the number density $n_{\rm H}(r)$, and the Toomre parameter $Q_{\rm T}(r)$, respectively.
Recall here that, in Figure \ref{fig_r-Mdot}, the evolutionary tracks of 
the disk outer radii $r_{\rm d}$ and 
instantaneous accretion rates $\dot{M}$ are indicated with 
curves with crosses.

First we see the zero-metallicity case (left column of Fig.\ref{fig_r-TQ}).
As the stellar mass increases,
the disk becomes larger since the infalling gas 
has larger angular momentum.
At more inner radius and with higher stellar mass,
the temperature becomes higher owing to the deeper gravitational potential
and thus higher viscous heating rate.
At the optically thick innermost region of $\la 10{\rm AU}$, 
the inefficient cooling makes the temperature very high 
and correspondingly the density low (top and middle panels).
Outside 10AU, the temperature remains about $300$ -- $3000{\rm K}$ 
by the ${\rm H}_2$-line cooling.
This disk temperature is similar to the envelope temperature, which is 
indicated with bars in top-left panel of Fig.\ref{fig_r-TQ} for each 
stellar-mass case.
In the bottom-left panel of Figure \ref{fig_r-TQ},
the Toomre parameter is almost constant at $Q_{\rm T}\simeq1$
in a large part of the disk by the self-regulation process due to gravitational torque.
In the outer disk of $\sim1000{\rm AU}$, however, 
the Toomre parameter falls below unity 
since the temperature is too low for this self-regulation mechanism 
to compensate the instability.
On the other hand, in the inner disk of a few AU,
the high temperature results in very high $Q_{\rm T}$ ($\gg 1$).
As indicated by the evolutionary track $(r_{\rm d},~\dot{M})$ 
in the panel for the $Z=0$ case of Figure \ref{fig_r-Mdot}, 
the accretion rate is always as high as $\ga10^{-3}\msun {\rm yr}^{-1}$
and the outer disk enters the unstable domain when $\ga10\msun$.

Next, we see the case of the disk with $10^{-4}\zsun$
(middle column of Fig.\ref{fig_r-TQ}).
In this case, the efficient dust cooling makes the temperature lower at $100$ -- $1000{\rm K}$ than the temperatures of the zero-metallicity disk.
Also, the temperature at the outer disk falls below the envelope temperature 
shown by the bars.
As a result, a large part of the disk becomes strongly unstable with $Q_{\rm T}=0.1$ -- $1$
in the low-mass regime of $1$ -- $10\msun$.
This can be seen also from the evolutionary track of $(r_{\rm d},~\dot{M})$
in the $10^{-4}\zsun$ panel of Figure \ref{fig_r-Mdot}.
At $\sim 0.1\msun$, the track enters into the domain of instability, 
which is extended by the dust cooling toward lower accretion rate. 
When $M_*>10\msun$, the decline of the accretion rate makes the disk 
stable again and the track exits from the domain of instability.
In addition, as seen in the case of $100\msun$, 
temperature at the outer region $\sim 1000{\rm AU}$ 
is higher than inside because the dust cooling is only efficient 
in the inner dense region for $M_* \ga 10\msun$.
In this fashion, the outer disk becomes stable for those cases.
In reality, however, 
the unstable disk with $Q_{\rm T}\sim 0.1$ would undergo 
catastrophic fragmentation before the stellar mass reaches $10\msun$.
Once fragmentation occurs and multiple protostars are formed inside 
a single pre-stellar core, the infalling material will be shared among them. 
With reduced accretion rate onto individual protostars, 
the circumstellar disks around them would be finally stabilized 
(Sec.\ref{sec_sf}).

Finally, we see the evolution of the $Z=\zsun$ disk
shown in the right column of Figure \ref{fig_r-TQ}.
The outer radius of the disk is two or three orders of magnitude larger 
than the zero-metallicity one at the same stellar mass, 
reflecting the more extended envelope (bottom panel of Fig.\ref{fig_envelope}).
The temperature is as low as $10{\rm K}$, 
similar to the envelope value.
This low disk temperature is due not only to efficient dust cooling but also to 
low heating rate because of the low accretion rate 
($\sim 10^{-6}\msun {\rm yr}^{-1}$).
In the inner disk of $\la10{\rm AU}$, notwithstanding,  
the temperature jumps up to $100$ -- $1000{\rm K}$
owing to the large optical depth and to the deep potential well.
As in the $Z=0$ case,
the Toomre parameter is self-regulated to $Q_{\rm T} \simeq 1-1.5$.
As seen in the panel for the $\zsun$ case in Figure \ref{fig_r-Mdot},
despite a larger domain of instability than in lower metallicity cases,
the low value of accretion rate prevent the evolutionary track from 
entering it. 
Although the disk is predicted to become unstable in our model 
for $M_{\ast} \ga 10\msun$, the stellar radiative heating
would boost the disk temperature from $ \sim 10$K 
in the case of such a massive star, 
and so our model would not be valid anymore.  
In addition, with accretion rate as low as $10^{-6}\msun {\rm yr}^{-1}$,
the star does not reach more massive than $10\msun$ within its lifetime $\la 10{\rm Myr}$, 
and also, radiative pressure on the accretion flow can be important in halting the accretion 
\citep{lar71,kah74,wol87}.

\begin{figure*}
\begin{center}
\includegraphics[width=170mm]{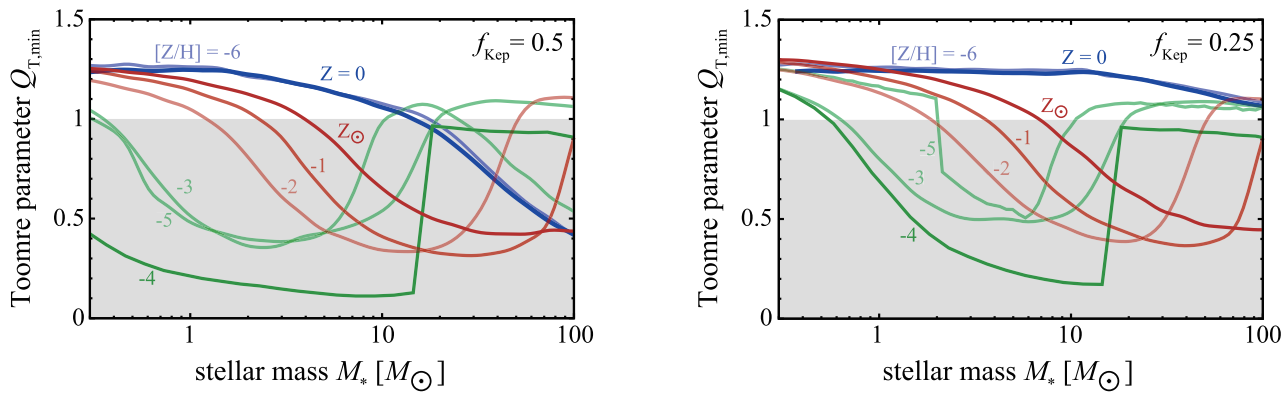}
\end{center}
\caption{The evolution of the minimum Toomre parameter $Q_{\rm T,min}$ in the disks 
with various metallicities 
$Z/\zsun = 0$, $10^{-6}$ (blue), 
$10^{-5},~10^{-4}, 10^{-3}$ (green), $10^{-2},~10^{-1},~{\rm and}~1$ (red) 
for the fiducial rotation parameter $f_{\rm Kep}=0.5$ ({\it left}), and 
a smaller value $f_{\rm Kep}=0.25$ ({\it right}).
In the panels, ${\rm [Z/H]}=\log(Z/\zsun)$.
The horizontal axes indicate the mass of the protostar $M_*$, which grows in time. 
}
\label{fig_MQ}
\end{figure*}

In Figure \ref{fig_MQ}, the minimum values of Toomre parameter $Q_{\rm T, min}$ in the disks are 
plotted as functions of the protostellar mass 
both for the fiducial rotation 
parameter $f_{\rm Kep}= 0.5$ ({\it left}) and for a smaller value $0.25$ ({\it right}).
First we see the fiducial case.
For the $Z=0$ and $10^{-6}\zsun$ disks, $Q_{\rm T,min}$ remains above unity 
until $M_{\ast}=20\msun$.
With metallicity of $10^{-5}-10^{-3}\zsun$,
the dust cooling becomes efficient enough to make the disks 
strongly unstable even in the early phase of $M_* \la 10\msun$.
In particular, the $10^{-4}\zsun$ disk is most unstable 
with $Q_{\rm T,min}\simeq 0.1$. 
The sudden stabilization in this case at $M_*\sim10\msun$, 
in which $Q_{\rm T,min}$ jumps up to $\simeq 1$, is caused by 
the decline of accretion rate as mentioned previously. 
With more metals $\ga10^{-2}\zsun$,
the disks are marginally stable $Q_{\rm T,min}\ga1$ 
for the stellar mass below several solar masses.
Next we see the case with slower rotation, $f_{\rm Kep}=0.25$.
The disks are now smaller than in the fiducial case (eq.\ref{eq_fkep}).
Due to higher temperature, disks are more stable at the inner radius. 
The smaller disks in the case of $f_{\rm Kep}=0.25$ are thus
more stable than those in the fiducial case.
However, the trend that the disks with two extreme values of metallicity, i.e., 
either around $Z=0$ or $Z=\zsun$, are marginally stable 
while those with metallicity in between 
$10^{-5}-10^{-3}\zsun$ are strongly unstable are common in both cases.

We here consider physical reason why disks with $10^{-5}$--$10^{-3}\zsun$ 
are strongly unstable
while those with higher and lower metallicity are relatively stable.
Suppose a disk with temperature of $T_{\rm disk}$ is formed as a result of
infall from an envelope with temperature $T_{\rm env}$. 
From equations (\ref{eq_q_cs_Mdot}) and (\ref{eq_infall}),
the Toomre parameter of the disk can be written as 
\begin{eqnarray}
	Q_{\rm T} \simeq \frac{3 \alpha c_{\rm s}^3}{G\dot{M}_{\rm env}}
	\simeq 3 \alpha \left ( \frac{T_{\rm disk}}{T_{\rm env}} \right)^{3/2}.
	\label{eq_T/T}
\end{eqnarray}
Now suppose that the temperatures in the disk and in the envelope are similar, 
$T_{\rm disk} \simeq T_{\rm env}$.
If the MRI always dominated the angular-momentum transfer 
($\alpha \ll 1$), the disk would become unstable with $Q_{\rm T} \ll 1$ 
from equation (\ref{eq_T/T}).
In reality, once $Q_{\rm T}$ approaches unity, the gravitational torque becomes 
efficient and $\alpha$ becomes close to unity. 
As a result, the disk is regulated to the marginally stable state of $Q_{\rm T}\ga1$.
On the other hand,
if the disk temperature is less than $\la 1/2$ of the envelope temperature,
the Toomre parameter is $Q_{\rm T}<1$ even with the efficient angular-momentum 
transport $\alpha=1$.
Such a low-temperature disk cannot avoid fragmention into pieces.
Similar phenomenon has been reported by \citet{kra10},
who studied the evolution of rapidly accreting protostellar disks by way of 
isothermal hydrodynamic simulations and demonstrated that binary systems 
are formed by disk fragmentation if the accretion rate onto the disks 
$\dot{M}_{\rm env}$ exceeds $3 c^3_{\rm s}/G$
\citep[see also][]{kim12}.
In extremely low-metallicity ($\la10^{-6}\zsun$) cases, 
the temperatures both in the envelope and in the disk have 
similar values of $\simeq 1000$K by the H$_2$ cooling. 
The protostellar disks are (marginally) stable in those cases.
With $10^{-5}$--$10^{-3}\zsun$, on the other hand, 
the dust cooling operates only in dense environments in disks 
(see eq.\ref{eq_ncr}) and the disk temperature is $\la100{\rm K}$, 
while the envelope temperature remains unaffected at several $100{\rm K}$.
This disk/envelope temperature difference renders the protostellar disks 
strongly unstable.
With metallicity exceeding $10^{-2}\zsun$,
the disks become marginally stable again
since the dust cooling is efficient both in the envelopes and disks.
In summary, the protostellar disks are most unstable in the metallicity range of
$10^{-5} \la Z/\zsun\la 10^{-3}$ because of the difference in dominant coolants 
in the disks and envelopes.

This conclusion could be appropriate independently of the maximum $\alpha$ value,
although our results shown in this section are obtained assuming $\alpha_{\rm GI,max}=1$.
In Figure \ref{fig_r-Mdot2},
we present the evolutionary tracks in $r_{\rm d}$-$\dot{M}$ planes for the maximum $\alpha$ of $0.07$ \citep{gam01,ric05}.
Since the Toomre parameter is inversely proportional to $\alpha$ (eq. \ref{eq_q_cs_Mdot} and \ref{eq_T/T}), disks are more unstable than those with $\alpha_{\rm GI,max}=1$.
However, we can also see that protostellar disks with $10^{-5}$ -- $10^{-3}\zsun$ are more unstable, $Q_{\rm T}\la0.1$ at $0.1\msun$, than zero and solar metallicity disks.
Although the choice of the maximum $\alpha$ value would alter the fragmentation boundary,
it does not change the conclusion that the most unstable metallicity is $10^{-5}\la Z/\zsun \la 10^{-3}$ (see also Sec. \ref{sec_uncertain} and App. \ref{sec_alpha}).

\section{Discussion} \label{sec_discussion}
\subsection{Star formation at different metallicities} \label{sec_sf}
Based on our results, we here discuss plausible scenario for star formation 
at different metallicities.
Our results indicate that star formation process with two extreme metallicities, i.e., 
at around zero- and the solar metallicity, is similar from the aspect of 
protostellar disk stability.
The first stars are formed in minihalos of $\sim 10^5\msun$ at redshift $z\sim30$,
from parent pre-stellar cores of $\sim 1000\msun$ formed by the $\rm H_2$-line-induced 
fragmentation, and grows with accretion rate as high as $10^{-3}\msunyr$
\citep[e.g.][]{brolar04,cia05,glo13}.
The photoionizing feedback from the protostars eventually 
dissipates the materials in the envelopes and disks, thereby shutting off
the accretion when the protostars reach $\sim 100\msun$
\citep{mck08,hos11,sta12,sus13,tan13,hir13}.
On the other hand, in our Galaxy with metallicity of about the solar value,
parent-core masses are typically $\sim 1\msun$,
and the protostellar accretion rates are $\sim 10^{-6}\msun{\rm yr}^{-1}$
\citep{shu87,mck07}.
While some envelope materials are expelled by magnetically driven protostellar outflow,
the majority of parent-core gas is expected to be converted to the newborn star \citep{mac12}.
In those ways,  
formation processes of the primordial and present-day stars look largely different
in terms of the mass-scale and typical accretion rate.
In both cases, nonetheless, 
the protostellar disks are self-regulated to a marginally stable state 
by the gravitational torque and would not go through catastrophic fragmentation 
(only modest one, if any),
and most of the infalling gas is channelled onto the primary or 
a small number of multiple stars.

In low-metallicity environments with $10^{-5}$--$10^{-3}\zsun$,
massive clumps of $100$--$1000\msun$ formed either by ${\rm H_2}$, 
HD ($\la 10^{-4}\zsun$) 
or metal-line ($\ga 10^{-4}\zsun$) cooling are expected to fragment again 
at higher density by the dust cooling to form subsolar-mass pre-stellar cores.
Some massive clumps, however, may fail to fragment in the dust cooling phase 
and survive due, for example, to lack of initial deformations or perturbations, 
or to small degree of rotation, etc.  
If so, two mass-scales of the pre-stellar cores, 
$\ga100\msun$ and $\la1\msun$, may exist simultaneously.
In the low-mass cores, all stars formed must be obviously low-mass.
Even in the massive cores, our results indicate that low-mass stars can
be formed by catastrophic disk fragmentation since protostellar disks with metallicity
in this range are strongly unstable with $Q_{\rm T}=0.1$ -- $1$ (Sec. \ref{sec_result2}).
The outcome would be a compact star-cluster, consisting of members of much lower-mass 
than the parent core.
It should be noted that
similar mechanism called the ``fragmentation induced starvation'' has been 
found to play a role in limiting the stellar mass
in the context of present-day massive star formation \citep{pet10}.
Our steady disk model cannot treat the evolution after the fragmentation, 
e.g., subsequent dynamical evolution of the resultant stellar system. 
We here try to estimate the mass and the number of stars formed by disk fragmentation.
Recall that the Toomre parameter is roughly proportional to the accretion rate 
(see eq. \ref{eq_q_cs_Mdot}).
In the case of $10^{-4}\zsun$, $Q_{\rm T}\simeq0.1$ for 
the accretion rate $\sim 10^{-4}-10^{-3}\msun {\rm yr}^{-1}$.
Therefore, once $\sim 10$ stars are formed as a result of the fragmentation 
and the accretion flow is divided equally among them, 
the disk around each star would become marginally stable.
On the other hand, the maximum stellar mass set by the photoionization feedback
is about $10\msun$ for this accretion rate \citep{hol94,ric97,tan13}.
Thus, even though the parent core is very massive $>100\msun$,
the end product would be a cluster of $\ga 10$ stars with mass less than  
$10\msun$.
This mass scale has, in fact, also been suggested by observations 
of carbon enhanced metal-poor stars, whose frequency are known to increase 
with decreasing metallicity. 
Their possible origin is secondary stars in binary systems, 
enriched with carbon by mass transfer from the primary stars, which 
might have already evolved to white dwarfs by now.  
In this scenario, 
the typical mass of the primaries is considered to be a few $\msun$ 
\citep{luc05,kom07,sud13}.
Our estimate above is concordance with this observational implication. 

Although this coincidence is encouraging, with our simple model, 
we can treat neither detailed fragmentation process nor the subsequent 
dynamical evolution of fragments.
Since typical distance between the fragments is as small as $100$ -- $1000{\rm AU}$,
some fragments may be scattered and/or merge with each other as a result of 
the gravitational interaction among them.
In fact, even in the case of the metal-free star formation, 
some (although not many) low-mass fragments are found to be ejected from the system 
as a consequence of multi-body gravitational interactions \citep{cla11,vor13}.
More vigorous disk fragmentations are expected in low-metallicity environments 
with $10^{-5}\la Z/\zsun \la 10^{-3}$, 
and thus a larger number of fragments would be ejected before growing massive.
This might be an origin of extremely metal-poor stars observed in the Galactic halo 
as well as of the free-floating planets.
For quantitative predictions about nonlinear physics such as 
the disk fragmentation and the multi-body gravitational interaction, 
sophisticated numerical hydrodynamics including chemical and radiative processes 
are awaited. 

The fragmentation of protostellar disks could produce very small objects with $\la0.01\msun$,
or gas-giant planets, if with little accretion thereafter.
In the metallicity range $Z \sim \zsun$, more fragmentation is expected in disks 
with lower metallicities from our results.
On the other hand, observations find
strong positive correlation between the discovery rate of giant exoplanets 
and the metallicity of host stars, with only few planets 
found in the range ${\rm [Fe/H]}<-0.5$
\citep{fis05,may11,mor13}.
This is not in contradiction to our results
as long as the dominant mechanism for giant-planet formation
is not the gravitational fragmentation of the disks
but the gas accretion onto the rocky cores.
The disk fragmentation, however, might have produced the planet discovered recently 
around a metal-poor star of ${\rm [Fe/H]}\simeq-2$ \citep{set12}
since large enough rocky cores to accrete gas are hard to form in 
such a low-metallicity environment \citep[e.g.,][]{joh12,joh13}.

\subsection{Uncertainties in our model} \label{sec_uncertain}
In the followings, we discuss uncertainties and effects not included in our model. 

The uncertainty in the maximum value of $\alpha$ is a significant problem to determine the fragmentation boundary.
We mainly adopted $\alpha_{\rm GI,max}=1$ as our reference value,
which is supported by numerical simulations of star formation processes \citep{kru07,kra10,cla11,kui11}.
On the other hand, the isolated disk simulations suggest a smaller maximum $\alpha$.
\citet{gam01} found that disks will fragment
if the disk cooling time $t_{\rm cool}\simeq \Sigma c_{\rm s}^2/\gamma(\gamma-1){\mathscr L}$ is shorter than $3\Omega_{\rm Kep}^{-1}$ using local simulation, where $\gamma$ is the adiabatic exponent.
Since the cooling time can be rewritten as $t_{\rm cool}\simeq4/9\gamma(\gamma-1)\alpha \Omega_{\rm Kep}$ (eq. 2, 4, 5, and 13) and \citet{gam01} assumed $\gamma=2$, this result corresponds to the maximum $\alpha$ value of $0.07$.
\citet{ric05} also supported $\alpha_{\rm max}=0.07$ in the cases of $\gamma=5/3$ and $7/5$ using the global simulation of isolated disks.
This disagreement between star formation simulations and isolated disk simulations may be due to the numerical resolution.
Recent higher resolution simulations of isolated disk demonstrated that fragmentation occurs even with much longer cooling timescale than $3\Omega_{\rm Kep}^{-1}$, indicating the smaller $\alpha_{\rm max}$ \citep{mer11,mer12,paa11}.
\citet{paa12} suggested the fragmentation is a stochastic process in the case of long cooling time and there are no certain boundary.
Otherwise, this disagreement may be related to the mass ratio $M_{\rm d}/M_*$.
Since the infall supplies a large amount of gas onto the protostellar disk in star formation process,
the disk mass keeps relatively high \citep{mac10,tsu11} and the efficient mass redistribution is needed.
Even in isolated disk simulations, \citet{lod05} showed that, if $M_{\rm d}/M_*\ga0.5$, the large-scale structure ($m=2$ mode) induces the strong mass redistribution of $\alpha>0.1$ without fragmentation until settling down to quasi-steady state.
In order to resolve this disagreement,
high resolution simulations of star formation with realistic cooling processes are required.
If the maximum value of $\alpha$ is as small as $0.07$ or less,
the contribution of MRI has also non-negligible effect.
The fixed $\alpha_{\rm MRI}$ of 0.01 is applied in this work.
In reality, however, MRI activity depends on ionization degree.
Especially, the existence of the dead zone where the MRI is not active by low ionization degree
would be important in studying the entire disk evolution \citep{gam96,san00,bai11}.
These uncertainties of $\alpha$ description affects the fragmentation boundary.
However, we would like to note that, even with this uncertainty, the protostellar disk at $10^{-5}\la Z/\zsun \la 10^{-3}$ is quite unstable because the disk is $Q_{\rm T} \ll 1$ even with large value of $\alpha_{\rm GI,max}=1$.
For the dependence of our results on the choice of $\alpha$, see also Appendix \ref{sec_alpha}.

In this work, we studied the disk structure using the steady mode ignoring temperature perturbation.
For the solar metallicity case, \citet{cos10} adopted a sophisticated fragmentation condition
considering the cooling-rate dependence on temperature perturbation.
The temperature perturbation leads some fraction of gas to have shorter cooling time than average, which accelerates fragmentation.
They found that, at the temperature regime of ice or dust sublimation,
the perturbation effect drops down the critical accretion rate for fragmentation
about an order of magnitude from the value simply estimated by the average field.
This effect may be important in low-metallicity star formation.
For example, in the metallicity of $10^{-4}\zsun$, the protostellar disk at $r\simeq1$--$10{\rm AU}$ has the dust sublimation temperature of about $1500{\rm K}$ (Fig. \ref{fig_r-TQ}).
Although the average field is $Q_{\rm T}\ga1$, fragmentation could be induced at this region by temperature perturbation.
Therefore, it is an important topic for future works to clarify the details of
the temperature-perturbation-induced instability in low-metallicity star formation.

We have adopted the same dust model as in the solar neighborhood.
However, the dust property in the early universe can be different.
In particular, the depletion factor $f_{\rm dep}$, i.e., the mass faction of metals 
in the dust phase, affects significantly the dust cooling rate \citep{sch12}.
The dust grains in the early universe are considered to be produced in supernovae,
and destructed by the reverse shocks.
The resultant depletion factor is typically a few percent, 
depending on the ambient density, much smaller than in the solar neighborhood 
($f_{\rm dep} \simeq 0.5$) \citep{noz07}.
Recent studies by \citet{noz12} and \citet{chi13}, however, showed that,
even though $f_{\rm dep}\ll 1$ initially, the dust grains could grow 
during the pre-stellar collapse owing to the sticking of silicate particles 
onto the dust, thereby boosting the depletion factor to the order of unity.
Since the protostellar disks are dense enough for the dust growth 
to work, our adoption of the Galactic value of $f_{\rm dep} \sim 1$ would be justified.

As the disk heating mechanism, we have considered only the viscous heating. 
Also the radiative heating form the central star, however, can be important
and stabilize the disks in some circumstances.
The degree of the stellar radiative heating depends on whether the innermost 
part of the disk is puffed up or not: if so, the outer disk is effectively 
shielded from the stellar radiation \citep[see][]{dul01}.
In fact, our result for $10\msun$ at $10^{-4}\zsun$ shows that the aspect ratio $H/r=0.2 {\rm~at~} 1{\rm AU}$, while $H/r=0.06 {\rm~at~} 100{\rm AU}$.
Although the innermost disk of $1{\rm AU}$ could be geometrically thick
due to high temperature there, we estimate the stellar heating without 
the shadowing as the conservative upper limit.
The equilibrium disk temperature $T_{\rm irr}$ determined by the balance 
between the stellar heating and the radiative cooling is 
\begin{eqnarray}
	T_{\rm irr} \simeq 74
	\left( \frac{M_*}{\msun} \right)^{-1/7}
	\left( \frac{L_*}{10^3\lsun} \right)^{2/7}
	\left( \frac{r}{100{\rm AU}} \right)^{-3/7}
	{\rm K},
\end{eqnarray}
where $L_*$ is the bolometric stellar luminosity \citep{kus70,chi97}. 
In evaluating $T_{\rm irr}$, we use $L_*$ from \citet{hos09}, who 
calculated the protostellar evolution for various metallicities.
For metallicity lower than $\sim 10^{-3}\zsun$, 
the stellar heating has little influence on the disk 
because the temperature determined by the viscous heating is already high with 
$100$ -- $1000{\rm K}$.
On the other hand, for $\ga10^{-2}\zsun$, 
the temperature by the viscous heating being only $\sim 10{\rm K}$, 
the stellar heating can be important unless the shadowing is effective 
\citep[see also][]{omu10}.
For the solar-metallicity disks, the importance of stabilization 
by the stellar radiation heating has already been pointed out by \citet{mat05,cai08}.
As seen in Section \ref{sec_result2},
the protostellar disks with $\ga10^{-2}\zsun$ are already relatively stable
only with the viscous heating and the additional heating by the stellar radiation 
will just make them even more stable.
We thus conclude that the stellar-heating stabilization 
does not qualitatively alter the general trend of disk-stability 
dependence on the metallicity.  

\section{Conclusion}
We have calculated the structure of protostellar disks with various metallicities 
by the steady-state $\alpha$-disk models, and examined their stability against 
self-gravity by the Toomre $Q_{\rm T}$ value. 
By constructing the envelope structure from the thermal evolution during 
the preceding collapse phase, the accretion rate onto the disk 
is evaluated as $\propto T_{\rm env}^{3/2}$, where 
$T_{\rm env}$ is the envelope temperature.
In this case, $Q_{\rm T}$ can be written using the ratio of the 
temperatures in the disk and in the envelope 
as $\sim \alpha_{\rm max}(T_{\rm disk}/T_{\rm env})^{3/2}$.
Here $\alpha_{\rm max}$ is the maximum value of the viscous parameter $\alpha$, which is about unity.
We have found that the protostellar disks can be classified into 
the following three metallicity regimes according to the disk stability, 
which is determined by the ratio $T_{\rm disk}/T_{\rm env}$.

(i) extremely low-metallicity ($\la10^{-6}\zsun$) disks: 
Both in the envelope and in the disk, 
the temperature is $\sim1000{\rm K}$ by the H$_2$-line cooling.
The disk is self-regulated to marginally stable state, 
$Q_{\rm T} \simeq 1$, by the gravitational torque. 

(ii) very low-metallicity ($10^{-5}$--$10^{-3}\zsun$) disks:
Temperature in the disk is $\sim 100$K by the dust cooling due to its high 
density $\ga 10^{10}{\rm cm}^{-1}$, while in the envelope it is several 100K by 
the H$_2$, HD ($\la 10^{-4} \zsun$) or metal-line cooling  ($\ga 10^{-4} \zsun$).  
This temperature difference results in a strongly unstable disk with $Q_{\rm T} \sim 0.1$, 
which is destined to fragment catastrophically.

(iii) metal-enriched ($\ga10^{-2}\zsun$) disks:
Temperatures are about $10{\rm K}$ both in the envelope and the disk  
by efficient dust cooling.
As in the extremely low-metallicity case, the disk is regulated to marginally stable 
state of $Q_{\rm T} \simeq 1$.

~

In the extremely metal-poor and metal-enriched environments, 
only the modest fragmentation of protostellar disks is expected. 
Therefore, most of the infalling material would accrete onto the primary or 
a small number of multiple stars. 
The typical mass of stars would be limited either by the available reservoir 
set by the parental core mass in the metal-enriched case
(at $\sim 0.1-1 \msun$) or by radiative feedbacks in the extremely metal-poor case 
(at $\sim 100 \msun$).

In the low-metallicity environment of $10^{-5}$--$10^{-3}\zsun$,
two mass-scales of parent cores may co-exist with $\la1\msun$ 
and $\ga100\msun$. The low-mass and high-mass cores are, respectively, 
formed by dust-cooling and line-cooling induced fragmentation during the 
pre-stellar collapse phase. 
Since the disks in this metallicity range are highly unstable, 
even in massive cores,  
low-mass stars would be formed by the protostellar disk fragmentation. 
Those stars would grow as massive as $\sim 10\msun$ or less, 
depending on complex interplay between multi-body gravitational interaction, 
gas accretion, and stellar radiative feedbacks.
This is roughly consistent with inferred mass-scale of metal-poor stars 
from observations of carbon-enhanced metal-poor stars 
in the Galactic halo \citep{luc05,kom07,sud13}. 

In conclusion, the fragmentation of the protostellar disks by 
the dust cooling, in cooperation with that during the pre-stellar collapse, 
accelerates the transition from massive to low-mass star formation mode
at the critical metallicity around $Z_{\rm crit} \simeq 10^{-5}\zsun$.

\section*{Acknowledgments}
The authors thank
Eduard Vorobyov,
Raffaella Schneider,
Masahiro N. Machida,
Takashi Hosokawa,
Satoshi Okuzumi,
Kengo Tomida,
Yusuke Tsukamoto,
Kohei Inayoshi
and Sanemichi Takahashi
for fruitful discussions and comments.
The authors also thank the anonymous referee for comments, which were useful to improve the original manuscript.
This work is supported by the Grants-in-Aid by the Ministry of Education, Science and Culture of Japan
(KO: 21684007, 25287040).

\appendix

\section{Critical density for thermal coupling of gas and dust} \label{sec_nc}
Here we find the threshold density $n_{\rm H, tc}$
above which the gas and dust are thermally coupled (eq. \ref{eq_ncr}).
As shown in Figure \ref{fig_Z-T_n_tau},
the dust temperature falls short of the gas temperature while dust cooling is not efficient.
Those temperatures become almost equal soon after the dust cooling dominates.

Suppose now that $T_{\rm d} \ll T$, i.e., before the thermal coupling,
 and the dust cooling is still inefficient.
In this case, the disks are optically thin and 
the radiation temperature $T_{\rm rad}$($\simeq \tau_{\rm P}T_{\rm d}$; 
eq. \ref {eq_Trad}) is even smaller than $T_{\rm d}$.
The dust temperature $T_{\rm d}$ is determined by the thermal equilibrium 
for the dust grains (eq. \ref{eq_dust_eq}).
Writing the mean free time for the gas-dust collision using the dust cross-section 
$\sigma_{\rm d}$ as 
\begin{equation}
	t_{\rm coll} = \left ({\sigma_{\rm d} \overline{n v}} \right)^{-1}, 
\end{equation}
where $\overline{n v}$ is the number densiy multiplied by velocity averaged over colliding
particles \citep{hol79,sch06,sch12}, 
the thermal equilibrim (eq. \ref{eq_dust_eq}) reads
\begin{eqnarray}
	4\sigma_{\rm SB} \rho \kappa_{\rm d,P} T_{\rm d}^4
	=
	2k_{\rm B} T n_{\rm d} \sigma_{\rm d} \overline{n v},
	\label{eq_dust_eq2}
\end{eqnarray}
where we have omitted the terms related to $T_{\rm rad}$ on the left-hand side 
and $T_{\rm d}$ on the right-hand side.
The opacity $\kappa_{\rm d,P}$ and the total cross-section of grains per unit gas mass 
$n_{\rm d}\sigma_{\rm d}/\rho$ are taken from the \citet{sem03}'s model 
and are approximately written as
\begin{eqnarray}
	&&\kappa_{\rm d,P}\simeq 7.4
	\left( \frac{Z}{\zsun} \right)
	\left( \frac{T_{\rm d}}{100{\rm K}} \right)^2
	{\rm~cm^{2}~g^{-1}}, \\
	&&n_{\rm d}\sigma_{\rm d}/\rho \simeq 470
	\left( \frac{Z}{\zsun} \right) {\rm~cm^{2}~g^{-1}},
\end{eqnarray}
with errors less than $20{\rm \%}$ for $T_{\rm d}=10-100{\rm K}$.
Substituting them into equation (\ref{eq_dust_eq2}), 
we obtain
\begin{eqnarray}
	T_{\rm d} \simeq 59
	\left( \frac{T}{100{\rm K}} \right)^{1/4}
	\left( \frac{n_{\rm H}}{10^{10}{\rm cm}^{-3}} \right)^{1/6}
	{\rm K}.
	\label{eq_Td}
\end{eqnarray}
The dust temperature $T_{\rm d}$ depends only explicitly 
on the gas temperature $T$ 
and density $n_{\rm H}$, but not on the metallicity $Z$.
Note, however, that the dependence on $Z$ comes through 
the gas temperature, which depends on $Z$.
Equating $T_{\rm d}$ with $T$ in equation (\ref{eq_Td}),
we can evaluate the density where the gas and dust thermally couple: 
\begin{eqnarray}
	n_{\rm H, tc} \simeq 2.3 \times 10^{11}
	\left( \frac{T}{100{\rm K}} \right)^{4.5}
	{\rm cm^{-3}}.
\end{eqnarray}
This value agrees well with our results and also results for 
collapsing cloud calculation by \citet{omu00}.

\section{Dependence on viscous parameter} \label{sec_alpha}
As we stated in Section \ref{sec_uncertain},
the prescription of $\alpha$ parameter has uncertainty. 
Therefore, we here demonstrate the robustness and sensitivity of our results to the value of $\alpha$.

\begin{figure}
\begin{center}
\includegraphics[width=75mm]{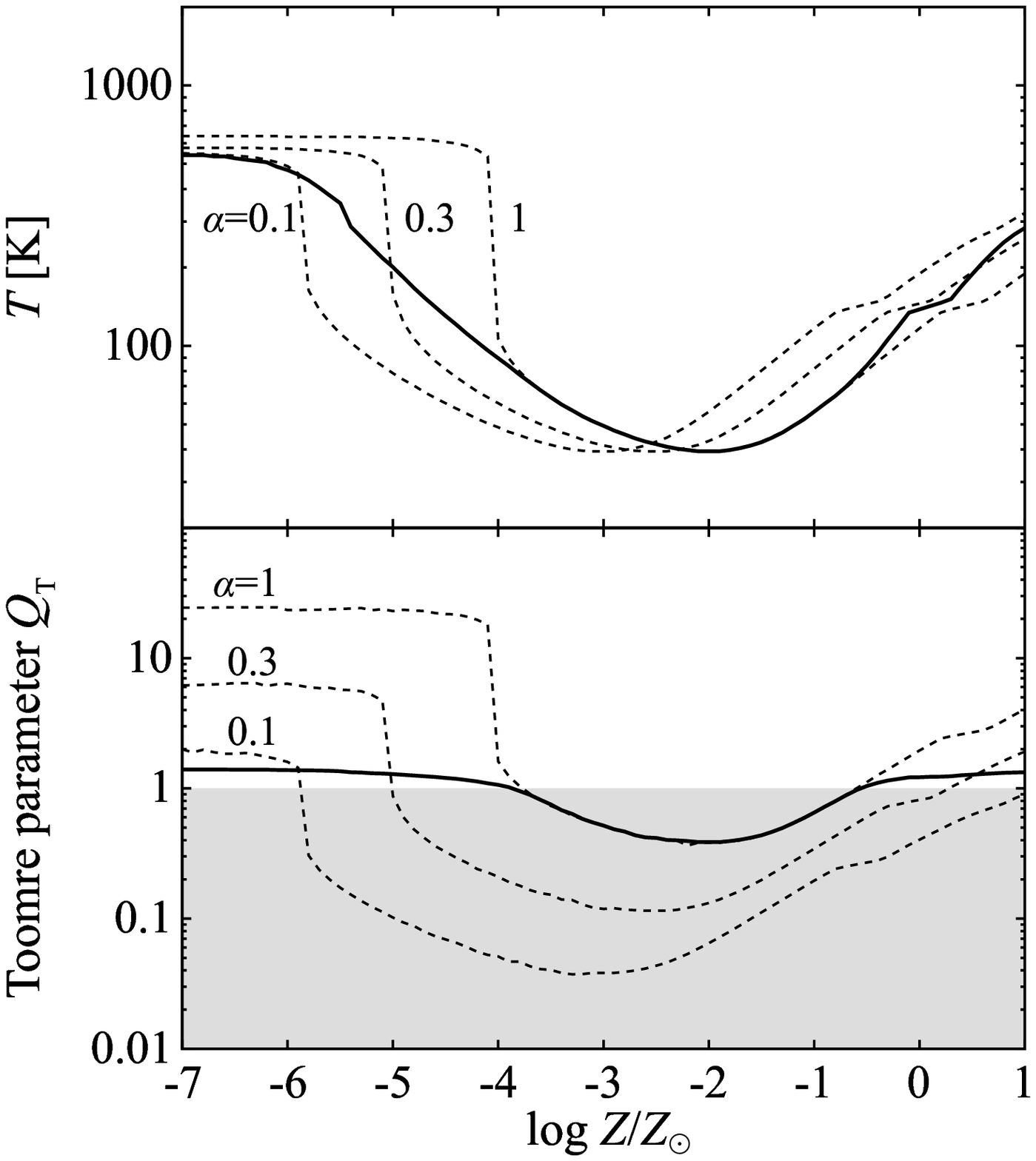}
\end{center}
\caption{
Comparison of the results with variable and constant $\alpha$'s.
The temperature ({\it top}) and the Toomre parameter ({\it bottom}) 
as functions of metallicity are shown 
for $M_*=10\msun$, $\dot{M}=10^{-4}\msunyr$, and $r=100{\rm AU}$.
{\it Solid:}
the case with the variable $\alpha$
$(= \exp \left(-{Q_{\rm T}^{10}}/{10} \right) + 10^{-2})$ used in the main part of 
this paper, 
{\it dashed:} the cases with constant $\alpha$ values
($\alpha=0.1,0.3,{\rm~and~} 1$).
}
\label{fig_alpha}
\end{figure}

Figure \ref{fig_alpha} shows the temperature $T$ ({\it top}) and the Toomre parameter $Q_{\rm T}$ ({\it bottom}) for models with our variable and three constant $\alpha$ values
(here we fix $M_*=10\msun$, $\dot{M}=10^{-4}\msunyr$, and $r=100{\rm AU}$).
In all cases, the metallicity dependence of temperature is qualitatively similar:
the temperature is lowest at $30{\rm K}$ around $10^{-3}$--$10^{-2}\zsun$,
while it is as high as several hundred kelvin at either higher or lower metallicities.
On the other hand, the value of Toomre parameter $Q_{\rm T}$ changes 
in response to the choice of the $\alpha$ value.
The $Q_{\rm T}$ value is approximately proportional to $\alpha$
in the constant $\alpha$ cases (see also eq. \ref{eq_q_cs_Mdot}), 
while, in the case of the variable $\alpha$,
it remains $Q_{\rm T} \simeq 1$ for a wide range of metallicity 
due to the regulation by the gravitational torque, which is effective 
only if $Q_{\rm T}\la1.5$.
Since the maximum $\alpha$ is set to unity in the variable $\alpha$ case,
the result in this case coincids with that with $\alpha=1$ when $Q_{\rm T}<1$.
Although the value of $Q_{\rm T}$ itself changes with the choice of $\alpha$,
the tendency that it is lowest around $10^{-3}$--$10^{-2}\zsun$ is common in all cases.
This fact confirms the robustness of our conclusion that the protostellar disks 
with $10^{-5}$--$10^{-3}\zsun$ are the most unstable.


\bsp

\label{lastpage}


\begin{thebibliography}{} 

\bibitem[Abel et al.(2002)]{abe02}
	Abel, T., Bryan, G.~L., \& Norman, M.~L.\ 2002, Science, 295, 93 

\bibitem[Bai(2011)]{bai11}
	Bai, X.-N.\ 2011, \apj, 739, 50 

\bibitem[Bate(1998)]{bat98}
	Bate, M.~R.\ 1998, \apjl, 508, L95

\bibitem[Bromm et al.(2001)]{bro01}
	Bromm, V., Ferrara, A., Coppi, P.~S., \& Larson, R.~B.\ 2001, \mnras, 328, 969

\bibitem[Bromm \& Larson(2004)]{brolar04}
	Bromm, V., \& Larson, R.~B.\ 2004, \araa, 42, 79 

\bibitem[Bromm \& Loeb(2004)]{bro04}
	Bromm, V., \& Loeb, A.\ 2004, \na, 9, 353 

\bibitem[Boley et al.(2006)]{bol06}
	Boley, A.~C., Mej{\'{\i}}a, A.~C., Durisen, R.~H., et al.\ 2006, \apj, 651, 517 
	
\bibitem[Boss(2002)]{bos02}
	Boss, A.~P.\ 2002, \apjl, 567, L149
	
\bibitem[Bromm \& Loeb(2003)]{bro03}
	Bromm, V., \& Loeb, A.\ 2003, \nat, 425, 812 

\bibitem[Cai et al.(2008)]{cai08}
	Cai, K., Durisen, R.~H., Boley, A.~C., Pickett, M.~K., \& Mej{\'{\i}}a, A.~C.\ 2008, \apj, 673, 1138 

\bibitem[Cai et al.(2006)]{cai06}
	Cai, K., Durisen, R.~H., Michael, S., et al.\ 2006, \apjl, 636, L149

\bibitem[Chabrier(2003)]{cha03}
	Chabrier, G.\ 2003, \pasp, 115, 763 

\bibitem[Chiaki et al.(2013)]{chi13}
	Chiaki, G., Nozawa, T., \& Yoshida, N.\ 2013, \apjl, 765, L3

\bibitem[Chiang \& Goldreich(1997)]{chi97}
	Chiang, E.~I., \& Goldreich, P.\ 1997, \apj, 490, 368 

\bibitem[Ciardi \& Ferrara(2005)]{cia05}
	Ciardi, B., \& Ferrara, A.\ 2005, \ssr, 116, 625 

\bibitem[Clark et al.(2008)]{cla08}
	Clark, P.~C., Glover, S.~C.~O., \& Klessen, R.~S.\ 2008, \apj, 672, 757 

\bibitem[Clark et al.(2011)]{cla11}
	Clark, P.~C., Glover, S.~C.~O., Smith, R.~J., et al.\ 2011, Science, 331, 1040 

\bibitem[Clarke(2009)]{cla09}
	Clarke, C.~J.\ 2009, \mnras, 396, 1066 

\bibitem[Cossins et al.(2009)]{cos09}
	Cossins, P., Lodato, G., \& Clarke, C.~J.\ 2009, \mnras, 393, 1157

\bibitem[Cossins et al.(2010)]{cos10}
	Cossins, P., Lodato, G., \& Clarke, C.\ 2010, \mnras, 401, 2587 

\bibitem[Davis et al.(2010)]{dav10}
	Davis, S.~W., Stone, J.~M., \& Pessah, M.~E.\ 2010, \apj, 713, 52

\bibitem[Dullemond et al.(2001)]{dul01}
	Dullemond, C.~P., Dominik, C., \& Natta, A.\ 2001, \apj, 560, 957

\bibitem[Dopcke et al.(2011)]{dop11}
	Dopcke, G., Glover, S.~C.~O., Clark, P.~C., \& Klessen, R.~S.\ 2011, \apjl, 729, L3 
	
\bibitem[Dopcke et al.(2013)]{dop13}
	Dopcke, G., Glover, S.~C.~O., Clark, P.~C., \& Klessen, R.~S.\ 2013, \apj, 766, 103 

\bibitem[Fischer \& Valenti(2005)]{fis05}
	Fischer, D.~A., \& Valenti, J.\ 2005, \apj, 622, 1102 

\bibitem[Gammie(1996)]{gam96}
	Gammie, C.~F.\ 1996, \apj, 457, 355 

\bibitem[Gammie(2001)]{gam01}
	Gammie, C.~F.\ 2001, \apj, 553, 174

\bibitem[Glover(2013)]{glo13}
	Glover, S.\ 2013, Astrophysics and Space Science Library, 396, 103

\bibitem[Hirano et al.(2013)]{hir13}
	Hirano, S., Hosokawa, T., Yoshida, N., et al.\ 2013, arXiv:1308.4456 

\bibitem[Hollenbach et al.(1994)]{hol94}
	Hollenbach, D., Johnstone, D., Lizano, S., \& Shu, F.\ 1994, \apj, 428, 654 (HJLS94)

\bibitem[Hollenbach \& McKee(1979)]{hol79}
	Hollenbach, D., \& McKee, C.~F.\ 1979, \apjs, 41, 555 

\bibitem[Hosokawa \& Omukai(2009)]{hos09}
	Hosokawa, T., \& Omukai, K.\ 2009, \apj, 703, 1810 

\bibitem[Hosokawa et al.(2011)]{hos11}
	Hosokawa, T., Omukai, K., Yoshida, N., \& Yorke, H.~W.\ 2011, Science, 334, 1250 

\bibitem[Johnson \& Li(2012)]{joh12}
	Johnson, J.~L., \& Li, H.\ 2012, \apj, 751, 81 

\bibitem[Johnson \& Li(2013)]{joh13}
	Johnson, J.~L., \& Li, H.\ 2013, \mnras, 431, 972 

\bibitem[Kahn(1974)]{kah74}
	Kahn, F.~D.\ 1974, \aap, 37, 149 

\bibitem[Kimura \& Tsuribe(2012)]{kim12}
	Kimura, S.~S., \& Tsuribe, T.\ 2012, \pasj, 64, 116 

\bibitem[Komiya et al.(2007)]{kom07}
	Komiya, Y., Suda, T., Minaguchi, H., et al.\ 2007, \apj, 658, 367 

\bibitem[Kratter et al.(2008)]{kra08}
	Kratter, K.~M., Matzner, C.~D., \& Krumholz, M.~R.\ 2008, \apj, 681, 375 

\bibitem[Kratter et al.(2010)]{kra10}
	Kratter, K.~M., Matzner, C.~D., Krumholz, M.~R., \& Klein, R.~I.\ 2010, \apj, 708, 1585 

\bibitem[Kroupa(2002)]{kro02}
	Kroupa, P.\ 2002, Science, 295, 82

\bibitem[Krumholz et al.(2009)]{kru09}
	Krumholz, M.~R., Klein, R.~I., McKee, C.~F., Offner, S.~S.~R., \& Cunningham, A.~J.\ 2009, Science, 323, 754 

\bibitem[Krumholz et al.(2007)]{kru07}
	Krumholz, M.~R., Stone, J.~M., \& Gardiner, T.~A.\ 2007, \apj, 671, 518 

\bibitem[Kuiper et al.(2011)]{kui11}
	Kuiper, R., Klahr, H., Beuther, H., \& Henning, T.\ 2011, \apj, 732, 20 

\bibitem[Kusaka et al.(1970)]{kus70}
	Kusaka, T., Nakano, T., \& Hayashi, C.\ 1970, Progress of Theoretical Physics, 44, 1580 

\bibitem[Larson(1969)]{lar69}
	Larson, R.~B.\ 1969, \mnras, 145, 271
	
\bibitem[Larson \& Starrfield(1971)]{lar71}
	Larson, R.~B., \& Starrfield, S.\ 1971, \aap, 13, 190 

\bibitem[Lin \& Pringle(1987)]{lin87}
	Lin, D.~N.~C., \& Pringle, J.~E.\ 1987, \mnras, 225, 607 

\bibitem[Lodato \& Rice(2004)]{lod04}
	Lodato, G., \& Rice, W.~K.~M.\ 2004, \mnras, 351, 630 

\bibitem[Lodato \& Rice(2005)]{lod05}
	Lodato, G., \& Rice, W.~K.~M.\ 2005, \mnras, 358, 1489 

\bibitem[Lucatello et al.(2005)]{luc05}
	Lucatello, S., Tsangarides, S., Beers, T.~C., et al.\ 2005, \apj, 625, 825 

\bibitem[Machida et al.(2010)]{mac10}
	Machida, M.~N., Inutsuka, S.-i., \& Matsumoto, T.\ 2010, \apj, 724, 1006 

\bibitem[Machida \& Matsumoto(2012)]{mac12}
	Machida, M.~N., \& Matsumoto, T.\ 2012, \mnras, 421, 588 

\bibitem[Matzner \& Levin(2005)]{mat05}
	Matzner, C.~D., \& Levin, Y.\ 2005, \apj, 628, 817 

\bibitem[Mayer \& Duschl(2005)]{may05}
	Mayer, M., \& Duschl, W.~J.\ 2005, \mnras, 358, 614

\bibitem[Mortier et al.(2013)]{mor13}
	Mortier, A., Santos, N.~C., Sousa, S., et al.\ 2013, \aap, 551, A112 

\bibitem[McKee \& Ostriker(2007)]{mck07}
	McKee, C.~F., \& Ostriker, E.~C.\ 2007, \araa, 45, 565 

\bibitem[McKee \& Tan(2008)]{mck08}
	McKee, C.~F., \& Tan, J.~C.\ 2008, \apj, 681, 771 

\bibitem[Meru \& Bate(2010)]{mer10}
	Meru, F., \& Bate, M.~R.\ 2010, \mnras, 406, 2279

\bibitem[Meru \& Bate(2011)]{mer11}
	Meru, F., \& Bate, M.~R.\ 2011, \mnras, 411, L1 

\bibitem[Meru \& Bate(2012)]{mer12}
	Meru, F., \& Bate, M.~R.\ 2012, \mnras, 427, 2022 

\bibitem[Mayor et al.(2011)]{may11}
	Mayor, M., Marmier, M., Lovis, C., et al.\ 2011, arXiv:1109.2497 

\bibitem[Nakamoto \& Nakagawa(1994)]{nak94}
	Nakamoto, T., \& Nakagawa, Y.\ 1994, \apj, 421, 640

\bibitem[Nakamoto \& Nakagawa(1995)]{nak95}
	Nakamoto, T., \& Nakagawa, Y.\ 1995, \apj, 445, 330

\bibitem[Nozawa et al.(2007)]{noz07}
	Nozawa, T., Kozasa, T., Habe, A., et al.\ 2007, \apj, 666, 955 

\bibitem[Nozawa et al.(2012)]{noz12}
	Nozawa, T., Kozasa, T., \& Nomoto, K.\ 2012, \apjl, 756, L35

\bibitem[Omukai(2000)]{omu00}
	Omukai, K.\ 2000, \apj, 534, 809  

\bibitem[Omukai(2001)]{omu01}
	Omukai, K.\ 2001, \apj, 546, 635

\bibitem[Omukai(2012)]{omu12}
	Omukai, K.\ 2012, \pasj, 64, 114 

\bibitem[Omukai \& Nishi(1998)]{omu98}
	Omukai, K., \& Nishi, R.\ 1998, \apj, 508, 141

\bibitem[Omukai et al.(2010)]{omu10}
	Omukai, K., Hosokawa, T., \& Yoshida, N.\ 2010, \apj, 722, 1793 

\bibitem[Omukai et al.(2005)]{omu05}
	Omukai, K., Tsuribe, T., Schneider, R., \& Ferrara, A.\ 2005, \apj, 626, 627 

\bibitem[Paardekooper(2012)]{paa12}
	Paardekooper, S.-J.\ 2012, \mnras, 421, 3286 

\bibitem[Paardekooper et al.(2011)]{paa11}
	Paardekooper, S.-J., Baruteau, C., \& Meru, F.\ 2011, \mnras, 416, L65 

\bibitem[Peters et al.(2010)]{pet10}
	Peters, T., Klessen, R.~S., Mac Low, M.-M., \& Banerjee, R.\ 2010, \apj, 725, 134

\bibitem[Pringle(1981)]{pri81} Pringle, J.~E.\ 1981, \araa, 19, 137 

\bibitem[Rice \& Armitage(2009)]{ric09}
	Rice, W.~K.~M., \& Armitage, P.~J.\ 2009, \mnras, 396, 2228 

\bibitem[Rice et al.(2003)]{ric03}
	Rice, W.~K.~M., Armitage, P.~J., Bate, M.~R., \& Bonnell, I.~A.\ 2003, \mnras, 339, 1025 

\bibitem[Rice et al.(2005)]{ric05}
	Rice, W.~K.~M., Lodato, G., \& Armitage, P.~J.\ 2005, \mnras, 364, L56 

\bibitem[Richling \& Yorke(1997)]{ric97}
	Richling, S., \& Yorke, H.~W.\ 1997, \aap, 327, 317 

\bibitem[Sano et al.(2000)]{san00}
	Sano, T., Miyama, S.~M., Umebayashi, T., \& Nakano, T.\ 2000, \apj, 543, 486 

\bibitem[Schneider et al.(2002)]{sch02}
	Schneider, R., Ferrara, A., Natarajan, P., \& Omukai, K.\ 2002, \apj, 571, 30 

\bibitem[Schneider et al.(2003)]{sch03}
	Schneider, R., Ferrara, A., Salvaterra, R., Omukai, K., \& Bromm, V.\ 2003, \nat, 422, 869 

\bibitem[Schneider et al.(2012)]{sch12}
	Schneider, R., Omukai, K., Bianchi, S., \& Valiante, R.\ 2012, \mnras, 419, 1566 

\bibitem[Schneider et al.(2006)]{sch06}
	Schneider, R., Omukai, K., Inoue, A.~K., \& Ferrara, A.\ 2006, \mnras, 369, 1437

\bibitem[Semenov et al.(2003)]{sem03}
	Semenov, D., Henning, T., Helling, C., Ilgner, M., \& Sedlmayr, E.\ 2003, \aap, 410, 611 

\bibitem[Setiawan et al.(2012)]{set12}
	Setiawan, J., Roccatagliata, V., Fedele, D., et al.\ 2012, \aap, 540, A141 

\bibitem[Shakura \& Sunyaev(1973)]{sha73}
	Shakura, N.~I., \& Sunyaev, R.~A.\ 1973, \aap, 24, 337 

\bibitem[Shi et al.(2010)]{shi10}
	Shi, J., Krolik, J.~H., \& Hirose, S.\ 2010, \apj, 708, 1716 

\bibitem[Shu(1977)]{shu77}
	Shu, F.~H.\ 1977, \apj, 214, 488 

\bibitem[Shu et al.(1987)]{shu87}
	Shu, F.~H., Adams, F.~C., \& Lizano, S.\ 1987, \araa, 25, 23 
	
\bibitem[Stacy \& Bromm(2013)]{sta13}
	Stacy, A., \& Bromm, V.\ 2013, \mnras, 1469 

\bibitem[Stacy et al.(2010)]{sta10}
	Stacy, A., Greif, T.~H., \& Bromm, V.\ 2010, \mnras, 403, 45 

\bibitem[Stacy et al.(2012)]{sta12}
	Stacy, A., Greif, T.~H., \& Bromm, V.\ 2012, \mnras, 422, 290 

\bibitem[Stahler et al.(1986)]{sta86}
	Stahler, S.~W., Palla, F., \& Salpeter, E.~E.\ 1986, \apj, 302, 590 

\bibitem[Stahler, Shu, \& Taam (1980)]{sta80}
	Stahler, S.~W., Shu, F.~H., \& Taam, R.~E.\ 1980, \apj, 241, 637 

\bibitem[Suda et al.(2013)]{sud13}
	Suda, T., Komiya, Y., Yamada, S., et al.\ 2013, \mnras, 432, L46 

\bibitem[Susa(2013)]{sus13}
	Susa, H.\ 2013, \apj, 773, 185 

\bibitem[Takahashi et al.(2013)]{tak13}
	Takahashi, S.~Z., Inutsuka, S.-i., \& Machida, M.~N.\ 2013, \apj, 770, 71 

\bibitem[Tan \& McKee(2004)]{tan04}
	Tan, J.~C., \& McKee, C.~F.\ 2004, \apj, 603, 383 

\bibitem[Tanaka \& Nakamoto(2011)]{tan11}
	Tanaka, K.~E.~I., \& Nakamoto, T.\ 2011, \apjl, 739, L50 

\bibitem[Tanaka et al.(2013)]{tan13}
	Tanaka, K.~E.~I., Nakamoto, T., \& Omukai, K.\ 2013, \apj, 773, 155

\bibitem[Toomre(1964)]{too94}
	Toomre, A.\ 1964, \apj, 139, 1217 

\bibitem[Tsukamoto \& Machida(2011)]{tsu11}
	Tsukamoto, Y., \& Machida, M.~N.\ 2011, \mnras, 416, 591

\bibitem[Vorobyov \& Basu(2010)]{vor10}
	Vorobyov, E.~I., \& Basu, S.\ 2010, \apj, 719, 1896 

\bibitem[Vorobyov et al.(2013)]{vor13}
	Vorobyov, E.~I., DeSouza, A.~L., \& Basu, S.\ 2013, \apj, 768, 131 

\bibitem[Walch et al.(2009)]{wal09}
	Walch, S., Burkert, A., Whitworth, A., Naab, T., \& Gritschneder, M.\ 2009, \mnras, 400, 13 

\bibitem[Wolfire \& Cassinelli(1987)]{wol87}
	Wolfire, M.~G., \& Cassinelli, J.~P.\ 1987, \apj, 319, 850 

\bibitem[Yoshida et al.(2006)]{yos06}
	Yoshida, N., Omukai, K., Hernquist, L., \& Abel, T.\ 2006, \apj, 652, 6 

\bibitem[Zhu et al.(2009)]{zhu09}
	Zhu, Z., Hartmann, L., Gammie, C., \& McKinney, J.~C.\ 2009, \apj, 701, 620 

\bibitem[Zinnecker \& Yorke(2007)]{zin07}
	Zinnecker, H., \& Yorke, H.~W.\ 2007, \araa, 45, 481 

\end{thebibliography}
\end{document}